\documentclass[pdftex,letterpaper,10pt]{article}

\pdfoutput=1
\usepackage[dvips]{graphicx}
\usepackage{mathptmx}
\usepackage{amsmath}
\usepackage{latexsym}
\usepackage{amssymb}
\usepackage{abstract}
\usepackage{enumerate}
\usepackage[dvips]{color}
\usepackage{titlesec}
\usepackage{abstract}
\usepackage[square,sort&compress]{natbib}
\usepackage{ctable}
\usepackage[top=1.5in, bottom=1.5in, left=1.5in, right=1.5in]{geometry}
\usepackage{color}
\usepackage{hyperref}
\hypersetup{colorlinks=false, urlcolor=blue, citecolor=green, linkcolor=red, citebordercolor={0 1 0}, linkbordercolor={1 0 0}}

\newcommand{\footnoteremember}[2]{\footnote{#2}\newcounter{#1}\setcounter{#1}{\value{footnote}}}
\newcommand{\footnoterecall}[1]{\footnotemark[\value{#1}]}

\begin{document}

\title{Electromagnetic waves and electron anisotropies downstream of supercritical interplanetary shocks}

\author{L.B. Wilson III\footnoteremember{1}{NASA Goddard Space Flight Center, Greenbelt, Maryland, USA.}, A. Koval\footnoteremember{5}{Goddard Planetary Heliophysics Institute, University of Maryland Baltimore County, Baltimore, Maryland, USA.}\footnoterecall{1}, A. Szabo\footnoterecall{1}, A. Breneman\footnoteremember{2}{School of Physics and Astronomy, University of Minnesota, Minneapolis, Minnesota, USA.}, C.A. Cattell\footnoterecall{2},\\  K. Goetz\footnoterecall{2}, P.J. Kellogg\footnoterecall{2}, K. Kersten\footnoterecall{2}, J.C. Kasper\footnoteremember{3}{Harvard-Smithsonian Center for Astrophysics, Harvard University, Cambridge, Massachusetts, USA.}, B.A. Maruca\footnoterecall{3}, M. Pulupa\footnoteremember{4}{Space Sciences Lab, University of California at Berkeley, Berkeley, California, USA.}}
\maketitle

\begin{abstract}
  We present waveform observations of electromagnetic lower hybrid and whistler waves with f${\scriptstyle_{ci}}$ $\ll$ f $<$ f${\scriptstyle_{ce}}$ downstream of four supercritical interplanetary (IP) shocks using the Wind search coil magnetometer.  The whistler waves were observed to have a weak positive correlation between $\delta$B and normalized heat flux magnitude and an inverse correlation with T${\scriptstyle_{eh}}$/T${\scriptstyle_{ec}}$.  All were observed simultaneous with electron distributions satisfying the whistler heat flux instability threshold and most with T${\scriptstyle_{\perp h}}$/T${\scriptstyle_{\parallel h}}$ $>$ 1.01.  Thus, the whistler mode waves appear to be driven by a heat flux instability and cause perpendicular heating of the halo electrons.  The lower hybrid waves show a much weaker correlation between $\delta$B and normalized heat flux magnitude and are often observed near magnetic field gradients.  A third type of event shows fluctuations consistent with a mixture of both lower hybrid and whistler mode waves.  These results suggest that whistler waves may indeed be regulating the electron heat flux and the halo temperature anisotropy, which is important for theories and simulations of electron distribution evolution from the sun to the earth.
\end{abstract}


\section{Introduction}  \label{sec:introduction}
\indent  Electromagnetic instabilities have been postulated to occur in and around collisionless shock waves for over 40 years \citep{tidman71a}.  A list of the waves/instabilities that are most commonly proposed to provide significant anomalous resistivity can be found in \citet{wu84a}.  Two types of waves have been of particular interest in collisionless shock energy dissipation:  lower hybrid and whistler waves.  Lower hybrid waves (LHWs) is a term used to encompass a broad range of waves and/or instabilities in plasma physics.  Though LHWs are often thought to exist as a simple linearly polarized electrostatic wave propagating perpendicular to the ambient magnetic field with f $\lesssim$ f${\scriptstyle_{lh}}$ ($=$ $\sqrt{ f{\scriptstyle_{ci}} f{\scriptstyle_{ce}} }$ for solar wind parameters), they can exist as a right-hand circularly polarized electromagnetic mode \citep{wu83a}.  Measurements at the terrestrial magnetopause have found that LHWs not only have k${\scriptstyle_{\parallel}}$ $\neq$ 0, the electromagnetic component can be comperable to the electrostatic one \citep{marsch83a, silin05a}.  \\
\indent  Whistler mode waves (WWs) can exist as a right-hand polarized electromagnetic mode propagating either parallel or at oblique angles to the magnetic field or as a quasi-electrostatic mode propagating at highly oblique angles with frequencies f${\scriptstyle_{ci}}$ $\leq$ f $\leq$ f${\scriptstyle_{ce}}$ \citep{brice64a}.  In the solar wind at low frequencies (f${\scriptstyle_{ci}}$ $\ll$ f $\leq$ f${\scriptstyle_{lh}}$), whistler mode waves can be very oblique electromagnetic modes which can lie on the same branch of the dispersion relation as LHWs \citep{marsch83a} and/or magnetosonic modes \citep{wu83a}.  Higher frequency (f${\scriptstyle_{lh}}$ $<$ f $\ll$ f${\scriptstyle_{ce}}$) WWs for which f $\rightarrow$ f${\scriptstyle_{ce}}$ have k${\scriptstyle_{\perp}}$ $\rightarrow$ 0 \citep{kennel66a} and can couple to multiple wave modes \citep{dyrud06}.  \\
\indent  Both LHWs and WWs can result from a number of different free energy sources.  WWs can be unstable to the heat flux carrying electrons \citep[\textit{e.g.}][]{gary94a, gary99a} and/or temperature anisotropy \citep[\textit{e.g.}][]{kennel66a} in the solar wind.  The relevant LHW free energy sources at collisionless shocks include, but are not limited to, currents \citep[\textit{e.g.}][]{lemons78a, matsukiyo06b}, ion velocity rings \citep[\textit{e.g.}][]{akimoto85c}, and/or heat flux carrying electrons \citep[\textit{e.g.}][]{marsch83a}.  More importantly, both modes can interact with electrons and ions and exchange energy between the two species.  It is well known that LHWs and highly oblique WWs can interact with high energy electrons parallel to the magnetic field and the bulk of the ion distributions perpendicular to the magnetic field through Landau resonance and nonlinear trapping \citep{wu83a, savoini95a}.  This is due to their large phase velocity parallel to the magnetic field ($\omega$/k${\scriptstyle_{\parallel}}$) and much smaller perpendicular phase velocity ($\omega$/k${\scriptstyle_{\perp}}$).  The stochastic energization of electrons and ions results in a nonthermal tail of the electrons parallel to the magnetic field and perpendicular ion heating.  Observation of these features in the ion and electron distributions in the ramp of one supercritical IP shock has been attributed to highly oblique whistler mode waves \citep{wilsoniii12d}.  \\
\indent  Although there have been numerous studies of LHWs in the magnetosphere  \citep[\textit{e.g.}][]{cattell95a}, there have only been a few observations done at the terrestrial bow shock \citep{wygant87, mellott88, walker08a} and only one observational study at an IP shock \citep{zhang98a}.  WWs have been observed in nearly every space plasma environment including, but not limited to, the terrestrial magnetosphere \citep[\textit{e.g.}][]{russell69a}, in the solar wind \citep[\textit{e.g.}][]{coroniti82a}, upstream of IP shocks \citep[\textit{e.g.}][]{russell83c, wilsoniii09a}, downstream of IP shocks \citep{moullard98a, moullard01}, upstream of planetary bow shocks \citep[\textit{e.g.}][]{hoppe81a}, and in the ramp of supercritical IP shocks \citep{wilsoniii12d}.  \\
\indent  Electron distributions in the solar wind are composed of a cold dense core (subscript c), a hotter more tenuous halo (subscript h), and a narrow field-aligned strahl \citep{feldman73b}.  Many theories \citep[\textit{e.g.}][]{gary94a, vocks05a, saito07} of the evolution of electrons from the Sun to the Earth suggest WWs scatter strahl electrons into the halo, which may explain the observed changes \citep[\textit{e.g.}][]{stverak09a} in the relative densities of strahl versus halo electrons.  Theories have also suggested that WWs constrain the heat flux carrying electrons \citep[\textit{e.g.}][]{gary94a, gary99a} and the halo/core electron temperature anisotropies \citep[\textit{e.g.}][]{vocks03a, saito08}.  The core electrons have been observed to isotropize due to binary collisions \citep[\textit{e.g.}][]{salem03}, while halo electrons appear to be constrained by the whistler anisotropy and fireshose instabilities in the slow solar wind \citep[\textit{e.g.}][]{stverak08a}.  These studies did not, however, examine waveform captures at the time of the particle distributions.  Thus, simultaneous observations of particle distributions with waveforms can test theories and improve our understanding of wave-particle dynamics in the solar wind.  \\
\indent  In this paper, we present an analysis of 47 individual waveform captures, sampled at 1875 samples per second, observed by the Wind search coil magnetometer downstream of four supercritical IP shocks.  The shocks were selected because they had search coil waveform captures with power peaked at f${\scriptstyle_{ci}}$ $\ll$ f${\scriptstyle_{sc}}$ $\leq$ f${\scriptstyle_{lh}}$, where f${\scriptstyle_{sc}}$ is the spacecraft frame frequency.  The particle distributions observed simultaneously satisfy the whistler heat flux and whistler anisotropy instability thresholds of \citet{gary94a} and \citet{gary99a}.  \\
\indent  The paper is organized as follows:  Section \ref{sec:data} introduces and outlines the data sets and analysis techniques; Section \ref{sec:observations} describes the waveform observations; Section \ref{sec:wavenumber} explains how we estimate wave numbers using multiple methods; Section \ref{sec:interpretation} discusses our analysis of possible free energy sources and interpretation of wave modes; and Section \ref{sec:conclusion} provides conclusions of our study.
\section{Data Sets and Analysis}  \label{sec:data}
\indent  Waveform captures were obtained from the Wind/WAVES instrument \citep{bougeret95a}, using the time domain sampler slow (TDSS) receiver, which provides a waveform capture (herein called TDSS event) of 2048 points with timespans of $\sim$1000 ms in this study with a lowest resolvable frequency of $\sim$3 Hz.  Therefore, waves with spacecraft frame frequencies below 3 Hz will not be observed.  For all the events in this study, TDSS events are composed of three magnetic ($\delta$B${\scriptstyle_{j}}$) and one electric ($\delta$E${\scriptstyle_{j}}$) field component, where $j$ $=$ x, y, or z.  We use $\delta$ to distinguish between the high frequency fields measured by the TDS instrument and the quasi-static magnetic fields measured by the fluxgate magnetometer.  Dynamic waveform analysis was performed through the use of the Morlet wavelet transform \citep{torrence98b} for each component of the TDSS events examined in this study.  For every TDSS wavelet presented herein, the spectra from all three components will have the same color scale range.  The Wind/WAVES instrument also contains an onboard time-averaged spectral intensity instrument, the thermal noise receiver (TNR), used to determine the local electron density (n${\scriptstyle_{e}}$) from the plasma line \citep{meyervernet89a}.  For more details about the WAVES instrument see \citep{bougeret95a}, and for analysis see \citet{wilsoniii10a}.  High time resolution magnetic field data was obtained from the dual triaxial fluxgate magnetometers \citep{lepping95}.  The time resolution is $\sim$0.092 s or $\sim$11 samples/second.  \\
\indent  We obtained full 4$\pi$ steradian low energy ($<$30 keV) electron and ion distributions from the Wind/3DP EESA and PESA particle detectors \citep{lin95a}.  Both EESA Low and PESA High can return full 4$\pi$ steradian distribution functions once every spin period ($\sim$3 s) in burst mode.  There was burst mode particle data for the 1998-08-26 and 2000-02-11 events but not the 1998-09-24 event.  The EESA Low detector measures electrons with $\lesssim$1.1 keV.  We use EESA Low (electrons) and PESA Low (ions) to calculate moments of the distribution functions and the anisotropy factor, A${\scriptstyle_{j}}$ $=$ T${\scriptstyle_{\perp,j}}$/T${\scriptstyle_{\parallel,j}}$ - 1, where $j$ corresponds to the jth-component of the electron distribution ($e$ $=$ entire distribution, $c$ $=$ core, and $h$ $=$ halo) and $\perp$($\parallel$) are defined with respect to ambient magnetic field, \textbf{B}${\scriptstyle_{o}}$.  The method for determining the break energy between halo and core electrons and the method to calculate the electron heat flux are outlined by \citet{wilsoniii09a}.  \\
\indent  Supplemental ion distributions were calculated from the two Faraday Cup (FC) ion instruments from the Wind SWE experiment \citep{ogilvie95}.  The SWE FCs can produce reduced ion distribution functions with up to 20 angular and 30 energy per charge bins every 92 seconds over an energy range of $\sim$150 eV to $\sim$8 keV ($\sim$1200 km/s proton) \citep{kasper06a}.  We will use the results of the reduced distributions produced by SWE to compare with our results from 3DP.  \\
\indent  The relevant shock parameters, determined by \citet{kasper}, for the events studied are shown in Table \ref{tab:machparams}, where $\hat{\textbf{n}}$ is the shock normal vector, V${\scriptstyle_{shn}}$ is the shock normal speed in the spacecraft (SC) frame, $\theta{\scriptstyle_{Bn}}$ is the angle between the shock normal vector and the upstream magnetic field vector, M${\scriptstyle_{f}}$ is the fast mode Mach number, and N${\scriptstyle_{i2}}$/N${\scriptstyle_{i1}}$ is the shock compression ratio.  The first critical Mach number, M${\scriptstyle_{cr}}$, was estimated for each IP shock using the methods outlined by \citet{edmiston84} assuming a polytrope index, $\gamma$ $=$ 5/3.  All of the shocks examined in this study were found to have M${\scriptstyle_{f}}$/M${\scriptstyle_{cr}}$ $>$ 1 (last column of Table \ref{tab:machparams}).  \\
\indent  The wave vector, \textbf{k}, and the polarization with respect to the ambient magnetic field were determined using Minimum Variance Analysis (MVA) \citep{khrabrov98}.  The low (f${\scriptstyle_{sc, low}}$) and high (f${\scriptstyle_{sc, high}}$) frequencies for each bandpass filter are determined from SC frame wavelet analysis.  Multiple standard Fourier bandpass filters were applied to each waveform and then we used MVA on specific subintervals.  The bandpass filters were applied to the entire TDSS event to reduce edge effects in the analysis of each subinterval.  Wave vector solutions are kept if $\lambda{\scriptstyle_{mid}}$/$\lambda{\scriptstyle_{min}}$ $\geq$ 10, where $\lambda{\scriptstyle_{j}}$ is the j$^{th}$ eigenvalue of spectral matrix with $\lambda{\scriptstyle_{max}}$ $>$ $\lambda{\scriptstyle_{mid}}$ $>$ $\lambda{\scriptstyle_{min}}$.  The details of this technique are discussed in \citet{wilsoniii09a}.  \\
\indent  The TDS data are displayed in two despun coordinate systems, GSE coordinates and field-aligned coordinates (FACs).  We define the FAC system with one axis parallel to \textbf{B}${\scriptstyle_{o}}$, a second parallel to \textbf{B}${\scriptstyle_{o}}$/$\mid$\textbf{B}${\scriptstyle_{o}}$$\mid$ $\times$ $\hat{\textbf{n}}$, and the third axis completes the right-handed system.  The long duration of the TDSS events with respect to spacecraft and variability in the magnetic field direction resulted in our using the instantaneous magnetic field interpolated from the fluxgate magnetometer measurements to rotate each vector in every TDSS event.
\section{Waveform Observations}  \label{sec:observations}
\indent  In Figure \ref{fig:3waveexamples} we show examples of each of the three waveform types to be examined in this study and summarize their physical properties.  We will explain how these properties were determined in more detail in the following section.  The three FAC components ($\delta$B${\scriptstyle_{\parallel}}$ (blue); $\delta$B${\scriptstyle_{\perp, 1}}$ (orange); and $\delta$B${\scriptstyle_{\perp, 2}}$ (green)), and a corresponding Morlet wavelet transform are plotted.  Overplotted on each wavelet transform are the lower hybrid (LH) resonance frequency, f${\scriptstyle_{lh}}$, and the electron cyclotron frequency, f${\scriptstyle_{ce}}$.  \\
\indent  Figure \ref{fig:3waveexamples}\textbf{A} is an example of an electromagnetic LHW.  The properties used to identify these modes in this study are:  (1) irregular waveform with power peaked at f${\scriptstyle_{sc}}$ $\lesssim$ f${\scriptstyle_{lh}}$; (2) mixtures of right and left-handed elliptical polarizations (SC Frame) with respect to the magnetic field; (3) typically propagate at angles greater than 45$^{\circ}$ from the magnetic field; (4) propagation direction shows no apparent dependence on solar wind direction; (5) have no apparent fluctuations for f${\scriptstyle_{sc}}$ $\gtrsim$ 60 Hz; (6) exhibit significant amplitudes in all three FAC components; and (7) typically have amplitudes $>$2 nT peak-to-peak.  There were 23/47 TDSS events classified as electromagnetic LHWs downstream of the 1998-08-26, 1998-09-24, and 2000-04-06 shocks.  \\
\indent  The WWs (Figure \ref{fig:3waveexamples}\textbf{B}) are characterized by six properties:  (1) peak power at f${\scriptstyle_{sc}}$ $\gtrsim$ 60 Hz; (2) smaller amplitudes than the electromagnetic LHWs ($\sim$1 nT peak-to-peak); (3) near circular right-hand polarization (SC Frame) with respect to the magnetic field; (4) only small $\delta$B${\scriptstyle_{\parallel}}$; (5) typically propagate within 30$^{\circ}$ of the magnetic field; and (6) typically propagate at $>$45$^{\circ}$ from the solar wind velocity.  There were 13/47 total TDSS events classified as WWs downstream of the 1998-08-26 and 2000-02-11 shocks.  \\
\indent  The third type of waveform observed in our study (Figure \ref{fig:3waveexamples}\textbf{C}) is a mixture of both LHWs and WWs.  These waves (MIXED) are characterized by five properties:  (1) the majority have large amplitudes ($>$2 nT peak-to-peak) for f${\scriptstyle_{sc}}$ $\lesssim$ f${\scriptstyle_{lh}}$ and smaller amplitudes ($\sim$1 nT peak-to-peak) for f${\scriptstyle_{sc}}$ $>$ f${\scriptstyle_{lh}}$; (2) right-hand polarizations (SC Frame) in both high and low frequencies; (3) significant amplitudes in $\delta$B${\scriptstyle_{\parallel}}$, $\delta$B${\scriptstyle_{\perp, 1}}$, and $\delta$B${\scriptstyle_{\perp, 2}}$ for low frequencies but only $\delta$B${\scriptstyle_{\perp, 1}}$ and $\delta$B${\scriptstyle_{\perp, 2}}$ for higher frequencies; (4) lower and higher frequency components show propagation characteristics consistent with the LHW and WWs, respectively; and (5) observed near sharp magnetic field gradients.  We observed 11/47 TDSS events classified as MIXED downstream of the 1998-09-24, 2000-02-11, and 2000-04-06 shocks.  \\
\indent  At first glance, these modes appear to be the simultaneous occurrence of LHWs with WWs superposed.  However, we will show in Section \ref{subsec:freeenergy} that the MIXED modes are observed under plasma conditions that are different enough from those found with the LHW and WWs to warrant a distinction.  In Section \ref{subsec:mvaresults}, we will define the separation between the high and low frequency components for the MIXED modes and discuss this in more detail.
\section{Wave Number and Rest Frame Frequency Estimates}  \label{sec:wavenumber}
\indent  In this section, we first discuss and summarize our minimum variance results in Section \ref{subsec:mvaresults}.  Then we discuss and summarize our estimates of the wave numbers and rest frame frequencies in Section \ref{subsec:dopplerresults}.
\subsection{Minimum Variance Results}  \label{subsec:mvaresults}
\indent  As discussed in Section \ref{sec:data}, we perform MVA on multiple frequency filters for multiple subintervals of each TDSS event.  Each subinterval yielding $\lambda{\scriptstyle_{mid}}$/$\lambda{\scriptstyle_{min}}$ $\geq$ 10 gives us an estimate of the wave vector ($\hat{\textbf{k}}$), which we can use to calculate wave propagation angles between the ambient magnetic field ($\theta{\scriptstyle_{kB}}$) and local \textbf{V}${\scriptstyle_{sw}}$ ($\theta{\scriptstyle_{kV}}$).  Note that for single point measurements with only magnetic fields, the sign of the wave vector cannot be definitively determined.  Therefore all angles are normalized to a 0$^{\circ}$ to 90$^{\circ}$ scale.  \\
\indent  Table \ref{tab:wavestats} presents a summary of the wave observations where the first column defines the wave type, the second column shows the number of TDSS events for each wave type, the third column shows the number of wave vectors satisfying the conditions discussed above, and the fourth column shows the mean plus or minus the standard deviation of the mean wave amplitudes.  Note that the wave amplitudes were determined from the filtered fields for each wave type.  \\
\indent  We examined 47 TDSS events and determined a total of 461 unique wave vectors from MVA.  $\sim$70$\%$ of the 118 LHW wave vectors satisfied $\theta{\scriptstyle_{kB}}$ $\geq$ 45$^{\circ}$ and $\sim$87$\%$ had $\theta{\scriptstyle_{kB}}$ $\geq$ 30$^{\circ}$.  The LHW wave vectors showed a broad range of $\theta{\scriptstyle_{kV}}$ peaked near 45$^{\circ}$.  Nearly all of the 138 WW wave vectors satisfied $\theta{\scriptstyle_{kB}}$ $\leq$ 45$^{\circ}$ and $\sim$83$\%$ had $\theta{\scriptstyle_{kB}}$ $\leq$ 30$^{\circ}$.  All of the WW wave vectors had $\theta{\scriptstyle_{kV}}$ $\geq$ 45$^{\circ}$ and $\sim$91$\%$ had $\theta{\scriptstyle_{kV}}$ $\geq$ 60$^{\circ}$.  \\
\indent  For the MIXED waves, the high and low frequency components were separated before wave normal angles were determined.  The TDSS events in our study all occurred under conditions with f${\scriptstyle_{lh}}$ $<$ 40 Hz.  In the SC frame, the power is primarily at frequencies $\lesssim$40 Hz and/or $\gtrsim$60 Hz.  We define high frequency (f${\scriptstyle_{high}}$) in the SC frame as (f${\scriptstyle_{sc}}$ $>$ f${\scriptstyle_{lh}}$) \textit{and} (f${\scriptstyle_{sc}}$ $>$ 40 Hz).  We define low frequency (f${\scriptstyle_{low}}$) in the SC frame as (f${\scriptstyle_{sc}}$ $\leq$ f${\scriptstyle_{lh}}$) \textit{or} (f${\scriptstyle_{sc}}$ $\leq$ 40 Hz).  We found 43/205 MIXED wave vectors satisfied f${\scriptstyle_{low}}$ and had a broad range of $\theta{\scriptstyle_{kB}}$ and $\theta{\scriptstyle_{kV}}$, both peaked near 45$^{\circ}$.  162/205 MIXED wave vectors satisfied f${\scriptstyle_{high}}$ and $\sim$67$\%$ had $\theta{\scriptstyle_{kB}}$ $\leq$ 45$^{\circ}$ and $\sim$78$\%$ had $\theta{\scriptstyle_{kV}}$ $\geq$ 45$^{\circ}$.  In summary, the MIXED modes satisfying f${\scriptstyle_{high}}$ have similar $\theta{\scriptstyle_{kB}}$ and $\theta{\scriptstyle_{kV}}$ as the WWs and those satisfying f${\scriptstyle_{low}}$ are similar to the LHWs.  While their propagation characteristics are consistent with the LHWs and WWs, they are observed under different plasma conditions than the LHWs(WWs).
\subsection{Doppler Shift Results}  \label{subsec:dopplerresults}
\indent  In the solar wind, waves can propagate at oblique angles with respect to the bulk flow (i.e. $\theta{\scriptstyle_{kV}}$ $\neq$ 90) causing a relatively stationary detector to observe a Doppler-shifted frequency.  In the spacecraft (SC) frame of reference, the wave frequency is a Doppler shifted signal given by Equation \ref{eq:doppler_0}.  To determine the rest frame frequency using single point measurements we need to assume a wave dispersion relation.  In this section, we will compare our wave number estimates from four different methods for one example (described in Appendix \ref{app:doppler}) and then discuss the results for all TDSS events analyzed.  \\
\indent  The results obtained for the waveform in Figure \ref{fig:3waveexamples}\textbf{C} calculation are shown in Table \ref{tab:wavenums}.  We used two frequency filters on this wave (f${\scriptstyle_{sc}}$ $\sim$ 5-30 Hz and $\sim$ 120-200 Hz) and so Table \ref{tab:wavenums} is separated into two sections.  The first part of each section of the table shows the background plasma parameters for this event along with the filter frequency, $\theta{\scriptstyle_{kB}}$, and $\theta{\scriptstyle_{kV}}$ ranges.  The second part of each section shows the wave number estimates.  \\
\indent  Table \ref{tab:wavenums} shows an increase in $\theta{\scriptstyle_{kV}}$ with increasing f${\scriptstyle_{sc}}$ for the example wave.  We found this to be a common trend with all the waves examined herein.  Although we do not have a definitive explanation why $\theta{\scriptstyle_{kV}}$ increases with increasing f${\scriptstyle_{sc}}$, it is possible that waves with small $\theta{\scriptstyle_{kV}}$ at higher frequencies are Doppler shifted above the Nyquist frequency of the search coil magnetometer ($\sim$940 Hz).  Note that the largest value of f${\scriptstyle_{sc}}$ $\sim$ 400 Hz, which is less than half the Nyquist frequency.  In our analysis, we use Equation \ref{eq:doppler_1} for both the WWs and MIXED modes satisfying f${\scriptstyle_{high}}$ and Doppler shifted Equation \ref{eq:lhws_em0} for LHWs and MIXED modes satisfying f${\scriptstyle_{low}}$.  \\
\indent  A summary of the results are shown in Table \ref{tab:waveparams} for kc/$\omega{\scriptstyle_{pe}}$, k$\rho{\scriptstyle_{ce}}$, k${\scriptstyle_{\parallel}}$$\rho{\scriptstyle_{ce}}$, k${\scriptstyle_{\perp}}$$\rho{\scriptstyle_{ce}}$, and $\omega$/$\Omega$.  Here, $\Omega$ $=$ $\omega{\scriptstyle_{lh}}$ ($=$ $\sqrt{ \Omega{\scriptstyle_{ci}} \Omega{\scriptstyle_{ce}} }$) for the LHWs and the MIXED modes satisfying f${\scriptstyle_{low}}$ and $\Omega$ $=$ $\Omega{\scriptstyle_{ce}}$ for the WWs and MIXED modes satisfying f${\scriptstyle_{high}}$.  For each wave type there are two rows, where the first row shows the absolute ranges and the second shows the mean plus or minus the standard deviation of the mean.  \\
\indent  The results for the LHWs and MIXED modes satisfying f${\scriptstyle_{low}}$ show a broad range of kc/$\omega{\scriptstyle_{pe}}$ and $\omega$/$\omega{\scriptstyle_{lh}}$.  The broad range of f${\scriptstyle_{sc}}$ may be due to Doppler broadening resulting from a range of $\theta{\scriptstyle_{kV}}$ for each TDSS event.  These results are consistent with theory \citep{marsch83a, wu83a}.  The broad range of $\omega$/$\Omega{\scriptstyle_{ce}}$ and kc/$\omega{\scriptstyle_{pe}}$ for the WWs and MIXED modes satisfying f${\scriptstyle_{high}}$ are consistent with theory \citep[\textit{e.g.}][]{kennel66a} and previous observations \citep{coroniti82a, lengyelfrey96, zhang98a} of whistler mode waves.  Thus, we argue that the observed modes are indeed on the whistler branch of the dispersion relation.  \\
\indent  The frequency spectrum, oblique propagation, and amplitudes of the LHWs are consistent with previous observations at collisionless shocks \citep{zhang98a, walker08a}.  The waveforms also show a resemblance to previous observations of a highly oblique class of waves, identified as whistler mode waves, observed downstream of the bow shock \citep{zhang98c}.  However, the LHWs presented herein are much larger in amplitude and observed much farther from the shock ramp.  The near circular right-hand polarization, amplitudes, and observed frequencies of the WWs are consistent with previous observations of narrow band whistler mode waves at IP shocks \citep{zhang98a, moullard98a, moullard01} and stream interaction regions \citep{breneman10a}.  At higher frequencies the wave vectors become more aligned with the magnetic field and more oblique to the solar wind velocity, consistent with theory \citep[\textit{e.g.}][]{kennel66a} and previous observations \citep{lengyelfrey96, zhang98a}.
\section{Free Energy and Interpretation of Waves}  \label{sec:interpretation}
\indent  In this section, we first introduce some previous work and theory on electromagnetic LHW and then WWs in the solar wind.  Following this, we will discuss the observed electron temperature anisotropies and normalized heat flux magnitude.  Finally we will discuss possible ion free energy sources.  \\
\subsection{Possible Free Energy Sources}  \label{subsec:freeenergy}
\begin{figure}[htbp]
  \begin{center}
    {\includegraphics[trim = 0mm 0mm 0mm 0mm, clip, width=16cm]{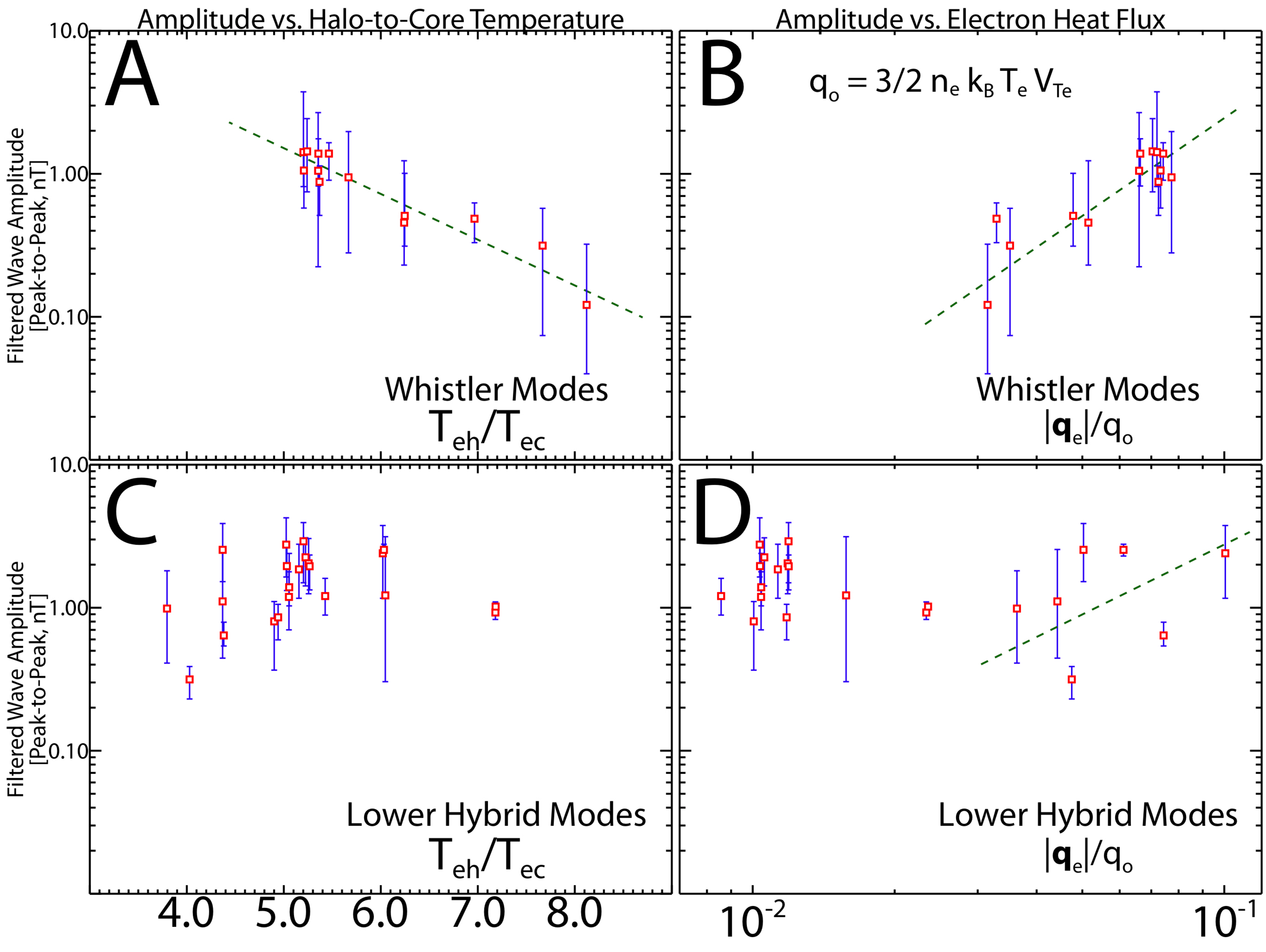}}
    \caption[Wave Amplitude Correlations]{Filtered wave amplitudes vs. T${\scriptstyle_{eh}}$/T${\scriptstyle_{ec}}$ and $\mid$$\vec{\textbf{q}}{\scriptstyle_{e}}$$\mid$/q${\scriptstyle_{o}}$ for the WW (panels \textbf{A} and \textbf{B}) and LHW (panels \textbf{C} and \textbf{D}) modes.  The red squares indicate the mean with the error bars showing the range of values for each TDSS event.  The heat flux magnitude is normalized by q${\scriptstyle_{o}}$, which is the free streaming heat flux saturation level \citep[\textit{e.g.}][]{salem03} (shown in panel \textbf{B}).  The green dashed lines are used to show qualitatively the trend for each set of points.}
    \label{fig:correlations}
  \end{center}
\end{figure}

\indent  There are multiple theories for the free energy sources for waves with f${\scriptstyle_{ci}}$ $\leq$ f $\leq$ f${\scriptstyle_{ce}}$ in the solar wind.  \citet{marsch83a, gary94a, gary99a} proposed that waves with f${\scriptstyle_{ci}}$ $\ll$ f $\leq$ f${\scriptstyle_{ce}}$ can extract free energy for wave growth from the heat flux carrying electrons in the solar wind.  Electromagnetic LHWs may also be driven by currents and/or gyrating ions \citep{akimoto85c}.  The theories for WW generation involve electron temperature anisotropies and the electron heat flux.  \citet{kennel66a} showed for a single bi-Maxwellian distribution that if:
\begin{equation}
  \label{eq:kennel_0}
  \frac{ \omega }{ \Omega{\scriptstyle_{ce}} } < \frac{ A{\scriptstyle_{e}} }{ A{\scriptstyle_{e}} + 1 }
\end{equation}
was satisfied, the distribution would be unstable to the whistler mode.  \citet{gary94a} showed that, even if no finite electron heat flux exists, if T${\scriptstyle_{\parallel h}}$/T${\scriptstyle_{\parallel c}}$ is small and T${\scriptstyle_{\perp h}}$/T${\scriptstyle_{\parallel h}}$ $>$ 1.01, then the whistler anisotropy instability may be excited.  Furthermore, \citet{gary99a} determined that when T${\scriptstyle_{\perp h}}$/T${\scriptstyle_{\parallel h}}$ $>$ 1.01 in the presence of a finite electron heat flux, the heat flux carrying electrons are always unstable to whistler heat flux instability.  \\
\indent  In an earlier study of whistler mode waves near IP shocks \citep{wilsoniii09a}, we found that the waves were observed simultaneously with particle distributions unstable to the whistler heat flux instability.  Based on estimates of the parameters of \citet{gary94a}, all of the WWs satisfied the threshold criteria for a whistler heat flux instability.  Furthermore, $\sim$77$\%$ (10/13) of the WWs were observed when T${\scriptstyle_{\perp h}}$/T${\scriptstyle_{\parallel h}}$ $>$ 1.01 and with a finite electron heat flux.  Interestingly, 19/23 of the electromagnetic LHWs were observed with T${\scriptstyle_{\perp h}}$/T${\scriptstyle_{\parallel h}}$ $>$ 1.01 and satisfied the same threshold criteria of \citet{gary94a} as the WWs.  For the MIXED modes, however, only 6/11 satisfied the threshold criteria of \citet{gary94a}.  \\
\indent  Figure \ref{fig:correlations} shows the filtered peak-to-peak wave amplitudes, $\delta$B, plotted vs. T${\scriptstyle_{eh}}$/T${\scriptstyle_{ec}}$ (panels \textbf{A} and \textbf{C}) and normalized heat flux magnitude (panels \textbf{B} and \textbf{D}), $\mid$$\vec{\textbf{q}}{\scriptstyle_{e}}$$\mid$/q${\scriptstyle_{o}}$, for the WW and LHWs.  Recall that for each TDSS event, we used multiple frequency filters and examined multiple subintervals, each with their own peak-to-peak $\delta$B.  Thus, the red squares indicate the mean values of $\delta$B and the error bars show the range of values for each TDSS event for all filters and subintervals.  We used green dashed lines to highlight qualitative trends in the data.  Note that the MIXED modes showed no relationship between $\delta$B and any of the parameters examined in our study (not shown).  \\
\indent  Figures \ref{fig:correlations}\textbf{B} and \ref{fig:correlations}\textbf{D} show a weak positive correlation between $\delta$B and $\mid$$\vec{\textbf{q}}{\scriptstyle_{e}}$$\mid$/q${\scriptstyle_{o}}$.  One can see that the correlation is much stronger for the WWs than the LHWs and that the relationship for the LHWs only exists for $\mid$$\vec{\textbf{q}}{\scriptstyle_{e}}$$\mid$/q${\scriptstyle_{o}}$ $\geq$ 0.03.  Figure \ref{fig:correlations}\textbf{A} shows a weak inverse relationship between $\delta$B and T${\scriptstyle_{eh}}$/T${\scriptstyle_{ec}}$ for the WWs, which may be explained by increased cyclotron damping.  The Landau and normal cyclotron resonant energies for the WWs range from $\sim$1-20 eV and $\sim$25-2100 eV, respectively, which supports our intrepretation of the relationship between $\delta$B and T${\scriptstyle_{eh}}$/T${\scriptstyle_{ec}}$.  There is no noticeable trend between $\delta$B and T${\scriptstyle_{eh}}$/T${\scriptstyle_{ec}}$ for the LHWs.  However, the LHWs were often observed near sharp changes in the background magnetic field, suggesting currents may provide some free energy for these modes.  \\
\indent  Previous simulation studies \citep[\textit{e.g.}][]{saito07} examining the effects of whistler mode turbulence on suprathermal electrons found that whistler mode waves were capable of scattering suprathermal electrons, imposing constraints on temperature anisotropies and strahl width and/or density.  They imposed wave amplitudes of $\delta$B/B${\scriptstyle_{o}}$ $\sim$ 0.07-0.10.  The WWs observed herein had $\delta$B/B${\scriptstyle_{o}}$ $\sim$ 0.01-0.16.  Therefore, the simulation results support our interpretation of the relationship between $\delta$B and $\mid$$\vec{\textbf{q}}{\scriptstyle_{e}}$$\mid$/q${\scriptstyle_{o}}$ and the inverse relationship between $\delta$B and T${\scriptstyle_{eh}}$/T${\scriptstyle_{ec}}$ for the WWs.  Furthermore, these waves are observed primarily when T${\scriptstyle_{\perp h}}$/T${\scriptstyle_{\parallel h}}$ $>$ 1.01 and $\mid$$\vec{\textbf{q}}{\scriptstyle_{e}}$$\mid$ $\neq$ 0, which are conditions suggested to always be unstable to the whistler heat flux instability \citep{gary99a}.  It is difficult to say whether the waves are caused by the observed distributions or whether the waves are producing the observed distributions.  Independent of the source of either effect, it is clear that the separation of the halo from the core is essential to understanding whistler related instabilities in the solar wind.
\subsection{Interpretation of Wave Modes}  \label{subsec:interpwaves}
\begin{figure}[htbp]
  \begin{center}
    {\includegraphics[trim = 0mm 0mm 0mm 0mm, clip, height=14cm]{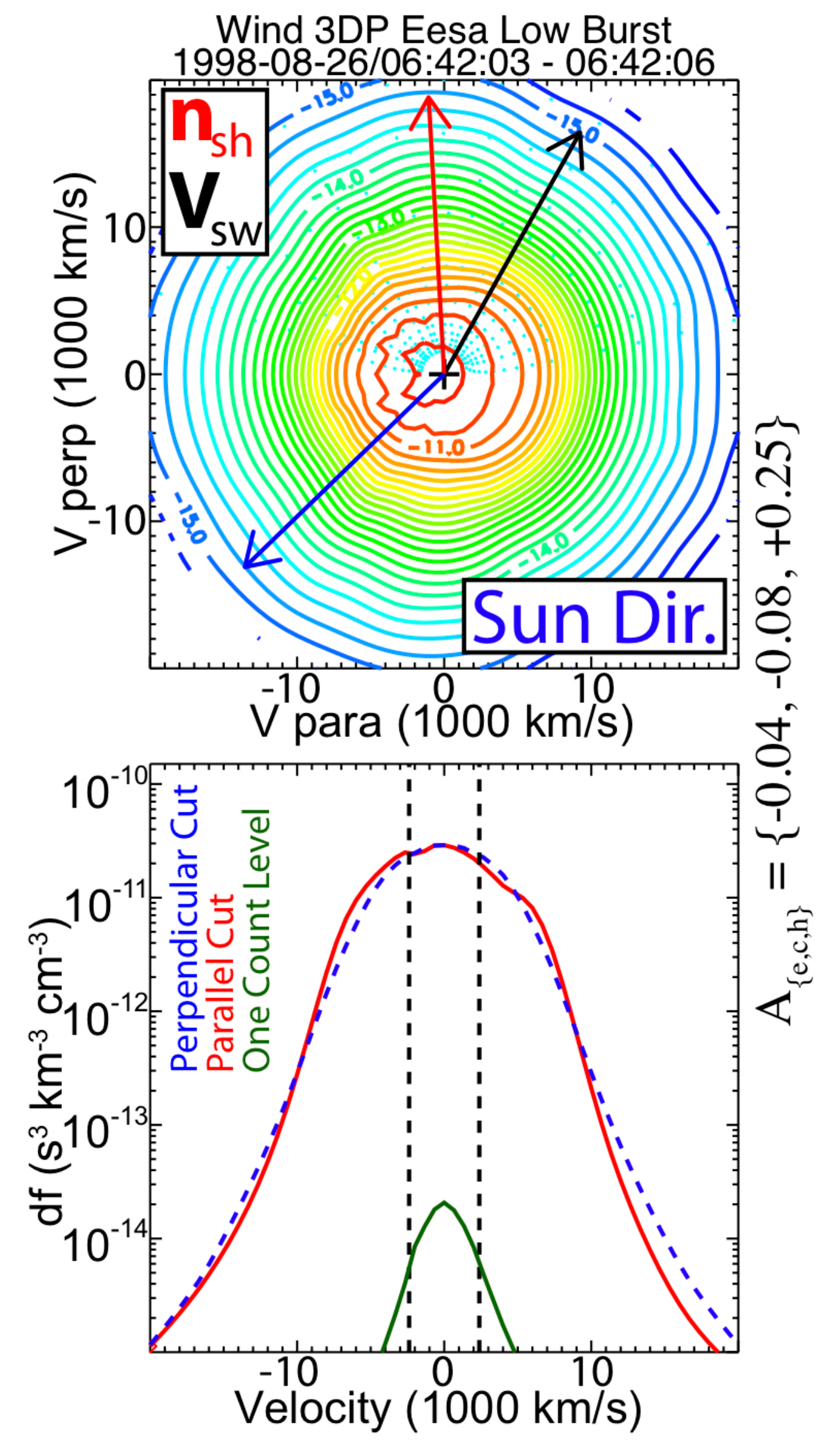}}
    \caption[Example Electron DF]{An example of an unstable electron distribution observed near a whistler mode wave downstream of the 1998-08-26 IP shock ramp.  The contour plots show contours of constant phase space density in the plane containing the ambient magnetic field (horizontal axis of contours) and solar wind velocity.  The shock normal direction (red arrow), solar wind velocity (black arrow), and sun direction (blue arrow) are projected onto the contour.  The bottom plot shows the parallel (solid red line) and perpendicular (dashed blue line) cuts of the distribution with associated one-count level (solid green line).  The vertical black dashed lines show the velocity corresponding to our estimate of the spacecraft potential.  To the right are values for the right-hand side of Equation \ref{eq:kennel_0} for the entire (e), core (c), and halo (h) component of the distribution.}
    \label{fig:particles}
  \end{center}
\end{figure}

\indent  The polarizations, rest frame frequencies, wave number range, observed proximity to magnetic field gradients, and relationship between $\delta$B and the $\mid$$\vec{\textbf{q}}{\scriptstyle_{e}}$$\mid$/q${\scriptstyle_{o}}$ are all consistent with the interpretation of the LHWs as electromagnetic lower hybrid waves.  The positive correlation between $\delta$B and $\mid$$\vec{\textbf{q}}{\scriptstyle_{e}}$$\mid$/q${\scriptstyle_{o}}$ may be due to stochastic electron acceleration of the halo electrons, since $\omega$/k${\scriptstyle_{\parallel}}$ $\gg$ $\omega$/k${\scriptstyle_{\perp}}$ for these modes \citep[\textit{e.g.}][]{cairns05a}.  However, below $\mid$$\vec{\textbf{q}}{\scriptstyle_{e}}$$\mid$/q${\scriptstyle_{o}}$ $\sim$ 0.03 we observe no relationship with $\delta$B.  Perhaps the waves are only affected by the heat flux carrying electrons above some threshold.  This may suggest that the LHWs, which are often observed near magnetic field gradients, are more dependent upon cross-field currents.  However, there is no one-to-one correlation between the two phenomena.  \\
\indent  The polarization, narrow band frequencies, nearly field-aligned propagation, and bursty waveform features are all consistent with the interpretation of the WWs as electromagnetic whistler mode waves.  Many of their properties are consistent with the narrow band electromagnetic whistler mode emissions called ``lion roars,'' originally observed in the magnetosheath in association with magnetic field depressions \citep[\textit{e.g.}][]{smith76a}.  \citet{zhang98c} observed lion roars in the terrestrial magnetosheath not associated with magnetic field depressions and interpreted their nearly unidirectional propagation as evidence that they propagated away from some source region.  \citet{masood06a} examined electron distributions looking for relationships between periods satsifying Equation \ref{eq:kennel_0} and observation of lion roars, yet they found no correlation.  \\
\indent  We believe that the reason previous studies \citep[\textit{e.g.}][]{masood06a} have failed to find a relationship between A${\scriptstyle_{e}}$ and lion roars (or whistler mode waves) is due to the use of single bi-Maxwellian electron distribution functions.  We find that only 5/13 WWs satisfy Equation \ref{eq:kennel_0} when using the entire distribution, however, that increases to 10/13 WWs when using only the halo.  We have already shown that all of the WWs satisfy the instability thresholds of \citet{gary94a}, which requires the separation of the core and the halo.  \\
\indent  Figure \ref{fig:particles} shows an example electron distribution that will highlight why we argue for the separation of the core and halo.  The values for A${\scriptstyle_{e,c,h}}$ are shown to the right of the figure.  Using these values, one can see that only the halo component would satisfy Equation \ref{eq:kennel_0}.  This example clearly illustrates why using moments of the entire distribution will not yield values that would predict an anisotropy instability.  This reiterates our emphasis on the treatment of the electron distribution using multiple components when looking for electron-related instability thresholds in the solar wind.  \\
\indent  Whether the WWs are lion roars is beyond the scope of this study.  We are interested in how they interact with the particles.  We do not observe any consistent relationship between the observation of WWs and changes in B${\scriptstyle_{o}}$ or n${\scriptstyle_{e}}$.  The correlation between $\delta$B and the $\mid$$\vec{\textbf{q}}{\scriptstyle_{e}}$$\mid$/q${\scriptstyle_{o}}$ and the fact that all the WWs were observed simultaneously with particle distributions unstable to the whistler heat flux instability suggest that the free energy for the WWs are the heat flux carrying electrons.  Recall that 10/13 WWs were observed with electron distributions whose halo satisfied Equation \ref{eq:kennel_0}.  In the previous section, we interpreted the inverse relationship between $\delta$B and T${\scriptstyle_{eh}}$/T${\scriptstyle_{ec}}$ as increased cyclotron damping.  Thus, we think the WWs may be causing T${\scriptstyle_{\perp h}}$/T${\scriptstyle_{\parallel h}}$ $>$ 1.01 and being driven by a whistler heat flux instability.  \\
\indent  The properties of the MIXED modes, at first glance, suggest that these are nothing more than the simultaneous observation of LHWs and WWs in the same TDSS event.  This can be seen when comparing LHWs and MIXED modes satisfying f${\scriptstyle_{low}}$ in Table \ref{tab:waveparams} and Figure \ref{fig:3waveexamples}.  The MIXED modes satisfying f${\scriptstyle_{high}}$ show nearly identical propagation characteristics to the WWs, as shown in Section \ref{subsec:mvaresults}.  However, only half of the MIXED modes were observed with electron distributions satisfying the instability thresholds of \citet{gary94a} and no relationship was observed between $\delta$B and the $\mid$$\vec{\textbf{q}}{\scriptstyle_{e}}$$\mid$/q${\scriptstyle_{o}}$ or T${\scriptstyle_{eh}}$/T${\scriptstyle_{ec}}$.  Thus, the MIXED modes do not appear to be interacting with the halo electrons in the same way as the WWs and LHWs.  \\
\indent  Since one can see in Figure \ref{fig:3waveexamples} that all the waves contain some fluctuations at f${\scriptstyle_{sc}}$ $<$ f${\scriptstyle_{lh}}$ and the wavelet transforms show power extending between the low and high frequency components, we examined effects arising from a coupling of the two modes.  We calculated estimates of threshold wave amplitudes for the modulational instability \citep[\textit{e.g.}][]{shapiro93a} and then converted the electric field amplitudes to magnetic field amplitudes \citep[\textit{e.g.}][]{huba78b}.  We find that 10/11 of the MIXED modes satisfy the threshold criteria for the modulational instability.  However, we also find that 18/22 LHWs satisfy the same criteria.  Therefore, it is not clear what drives the MIXED modes or the LHWs.
\section{Discussion and Conclusions}  \label{sec:conclusion}
\indent  We present the properties of electromagnetic lower hybrid and whistler mode waves downstream of supercritical IP shocks.  The work presented herein can be summarized by the following points:  \\
{\noindent \bf 1.}  The lower hybrid wave properties are consistent with theory \citep{marsch83a, wu83a} and previous observations at collisionless shocks \citep{zhang98a, walker08a}.  These modes are often observed near magnetic field gradients and they show a very weak positive correlation between $\delta$B and the $\mid$$\vec{\textbf{q}}{\scriptstyle_{e}}$$\mid$/q${\scriptstyle_{o}}$.  However, it is unclear what the source of free energy is for these modes;  \\
{\noindent \bf 2.}  The whistler mode wave properties are consistent with theory \citep[\textit{e.g.}][]{kennel66a} and previous observations \citep{coroniti82a, lengyelfrey96, zhang98a}.  We observe a weak positive correlation between $\delta$B and $\mid$$\vec{\textbf{q}}{\scriptstyle_{e}}$$\mid$/q${\scriptstyle_{o}}$ and an inverse relationship between $\delta$B and T${\scriptstyle_{eh}}$/T${\scriptstyle_{ec}}$.  All of these modes are observed with electron distributions that satisfy the whistler heat flux and whistler anisotropy instability thresholds \citep{gary94a, gary99a};  \\
{\noindent \bf 3.}  The events showing mixed wave features may be due to a modulational instability or they may be coincidental observations (i.e. merely simultaneous temporal observations).  They show no relationship between $\delta$B and any of the quantities examined for the other two modes.  Thus, the source of free energy for these modes is unclear;  \\
{\noindent \bf 4.}  The separation of the core and halo is absolutely necessary for instability analysis in the solar wind.  This was clearly demonstrated with our example distribution in Section \ref{subsec:interpwaves}; and  \\
{\noindent \bf 5.}  The waves presented herein have amplitudes as large as $\delta$B/B${\scriptstyle_{o}}$ $\sim$ 0.3, rest frame frequencies from $<$ $\omega{\scriptstyle_{lh}}$ to $\sim$ 0.9$\Omega{\scriptstyle_{ce}}$, and k$\rho{\scriptstyle_{ce}}$ $\sim$ 0.01-5.0.  \\
\indent  The observations suggest that the whistler mode waves are receiving free energy from the heat flux carrying electrons and producing T${\scriptstyle_{\perp h}}$/T${\scriptstyle_{\parallel h}}$ $>$ 1.01.  The lower hybrid waves may be interacting with the heat flux carrying electrons but only for sufficiently large values of $\mid$$\vec{\textbf{q}}{\scriptstyle_{e}}$$\mid$/q${\scriptstyle_{o}}$.  Since lower hybrid waves have $\omega$/k${\scriptstyle_{\parallel}}$ $\sim$ V${\scriptstyle_{T_{e}}}$ $\gg$ $\omega$/k${\scriptstyle_{\perp}}$, it may be that smaller heat fluxes do not have noticeable effects on the wave amplitudes.  At larger values, perhaps these electrons add to the wave amplitude or perhaps the larger amplitude waves add to the heat flux.  Their proximity to magnetic field gradients would be consistent with cross-field current instabilities or gradient-drift instabilities, but not all are observed near magnetic field gradients.  We also observe waves with high and low frequency fluctuations independently consistent with whistler and lower hybrid waves, respectively.  These events have no dependence of $\delta$B on any of the parameters examined in our study.  Therefore, the only conclusive argument we can make is that the whistler mode waves are driven by the heat flux carrying electrons and cause perpendicular heating of the halo through cyclotron interactions.  \\
\indent  Though we cannot determine whether the waves cause the observed distributions or vice versa, we argue that the wave amplitudes clearly interact with the halo electrons.  These results have implications for theories of the evolution of particle distributions from the Sun to the Earth.  For instance, the observation of whistler mode waves with unstable electron distributions supports theories suggesting whistler mode waves limit the halo electron temperature anisotropies and solar wind strahl.  Note that the observed whistler amplitudes can exceed the estimates used in the simulations \citep[\textit{e.g.}][]{saito07} that have led to these theories.  These results suggest that the evolution of the halo/strahl from the Sun to the Earth could be explained by wave-particle interactions with whistler mode waves.  However, there is another factor which has not been mentioned.  The observed waves and distributions all occur downstream of strong IP shocks.  This leads to two possibilities about the evolution of the electron distributions:  (1)  the modifications are due to transient events like IP shocks or (2) the wave-particle scattering suggested above is a global process but is enhanced downstream of IP shocks.  Further investigation during quiet solar wind periods is necessary to resolve this issue.  Future work will compare periods of fast and slow solar wind streams.

\section{Acknowledgments}  \label{sec:acknowledgments}
  \indent  We thank S.D. Bale, J.R. Wygant, and R. Lysak for useful discussions of the physics involved in our study.  All Wind spacecraft data were produced under Wind MO\&DA grants.  This research was partially supported by NESSF grant NNX07AU72H, grant NNX07AI05G, the Dr. Leonard Burlaga/Arctowski Medal Fellowship, and UCB work sponsored by NASA grant NNX10AT09G.

\appendix 
\section{Doppler Shift Calculations} \label{app:doppler}
\begin{figure}[htb]
 \begin{center}
   {\includegraphics[trim = 0mm 0mm 0mm 0mm, clip, width=16cm]{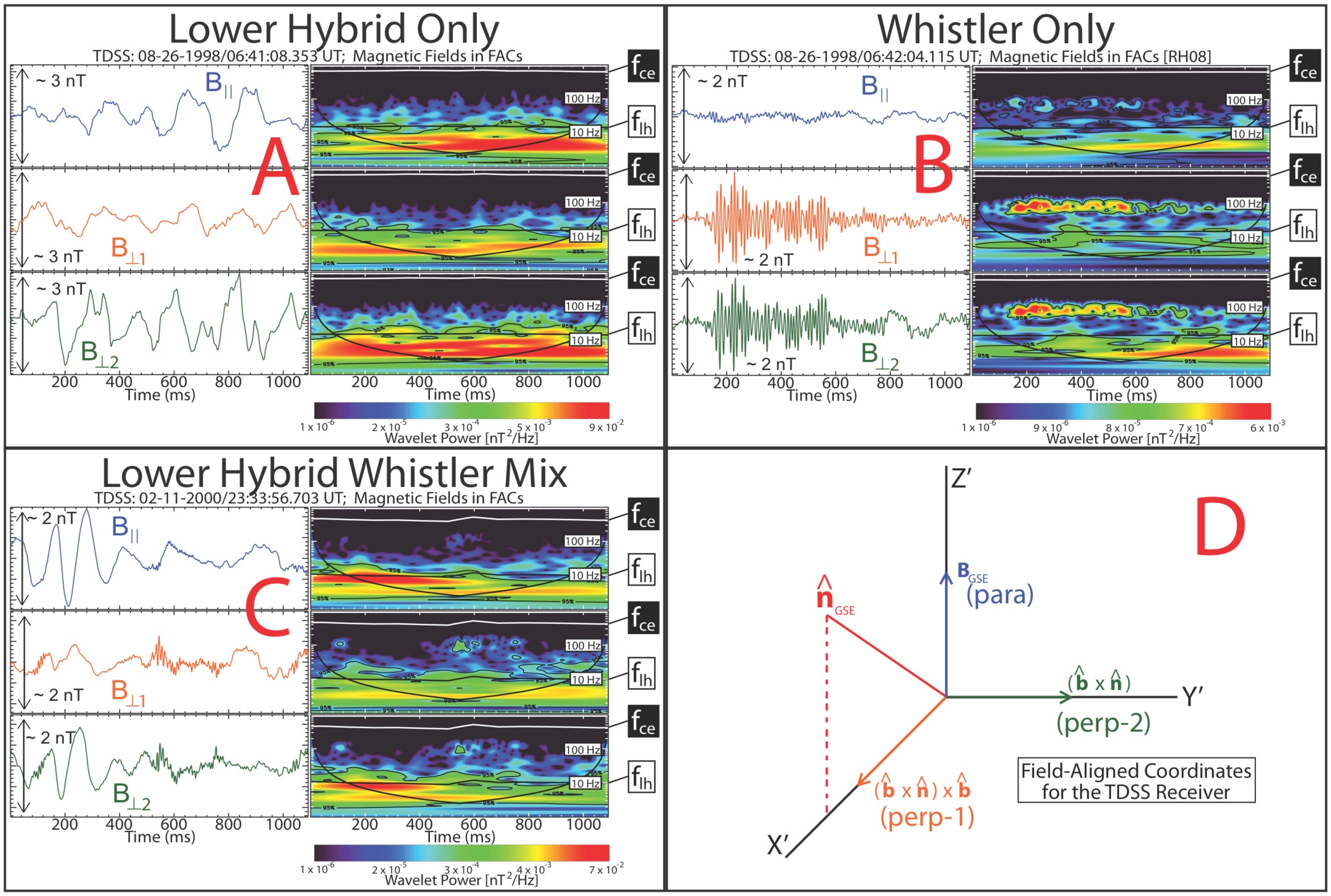}}
   \caption[3 Wave Type Examples with wavelets]{Three examples representative of the three wave types observed in this study including the LHW (panel \textbf{A}), WW (panel \textbf{B}), and MIXED (panel \textbf{C}) modes.  Panel \textbf{D} shows the right-handed coordinate system used to rotate the waveforms.  The left-hand column of each wave event panel (\textbf{A}-\textbf{C}) shows the FACs of the magnetic field and the right-hand corresponding column the wavelet spectrogram.  The relative amplitudes of the waveforms are illustrated by the vertical black arrows.  In the wavelet panels two lines mark the lower-hybrid resonance (black line) and the electron cyclotron frequency (white line).  The wavelet spectrograms are uniformly scaled for each individual waveform in nT$^{2}$/Hz.  On each wavelet spectrogram are a bowl-like line and contours marking the cone of influence and 95$\%$ confidence levels, respectively \citep{torrence98a}.}
   \label{fig:3waveexamples}
 \end{center}
\end{figure}

\indent  In this section, we define the four methods used to estimate ranges for the wave numbers, $\mid$\textbf{k}$\mid$, of the observed wave modes.  Since we are limited to single point measurements, we must make some assumptions about the fundamental properties of the observed modes.  Below, we will introduce the theory and describe how we estimate values for $\mid$\textbf{k}$\mid$.  \\
\indent  In the solar wind, $\Omega{\scriptstyle_{ce}}$ $\ll$ $\omega{\scriptstyle_{pe}}$ (where $\omega{\scriptstyle_{ps}}$ is the plasma frequency and $\Omega{\scriptstyle_{cs}}$ is the cyclotron frequency of species s) which limits the number of allowable normal modes of propagation with $\Omega{\scriptstyle_{ci}}$ $<$ $\omega$ $\leq$ $\Omega{\scriptstyle_{ce}}$.  The three most commonly cited wave modes in this frequency range in the solar wind are the magnetosonic \citep{kraussvarban91}, lower hybrid \citep{marsch83a}, and whistler mode waves \citep{coroniti82a}.  At low frequencies (f${\scriptstyle_{ci}}$ $<$ f $\leq$ f${\scriptstyle_{lh}}$), all three modes can lie on the same index of refraction surface \citep[\textit{e.g.}][]{wu83a}.  Therefore, let us start with the whistler mode.  The cold plasma index of refraction \citep{gurnett05} for an obliquely propagating whistler mode wave and $\omega^{2}$ $\ll$ $\omega{\scriptstyle_{pe}}^{2}$ is given by:
\begin{equation}
  \label{eq:indexref1}
  n^{2} = \frac{k^{2} c^{2}}{\omega^{2}} = \frac{ \omega{\scriptstyle_{pe}}^{2} }{\omega (\Omega{\scriptstyle_{ce}} \cos{ \theta{\scriptstyle_{kB}} } - \omega) }
\end{equation}
where $\omega$ is the rest frame angular frequency and $\theta{\scriptstyle_{kB}}$ is the angle between the wave vector and the ambient magnetic field.  Solving Equation \ref{eq:indexref1} for $\mid$\textbf{k}$\mid$ will give us our first wave number estimates.  \\
\indent  In the solar wind, the plasma flows with a velocity \textbf{V}${\scriptstyle_{sw}}$ relative to the spacecraft frame of reference (ignoring spacecraft motion), which causes any observed frequency to be Doppler shifted by:
\begin{equation}
  \label{eq:doppler_0}
  \omega{\scriptstyle_{sc}} = \omega + \textbf{k} \cdot \textbf{V}{\scriptstyle_{sw}}
\end{equation}
where $\omega{\scriptstyle_{sc}}$ is the spacecraft frame or observed frequency.  We can solve Equation \ref{eq:indexref1} for $\omega$ and substitute into Equation \ref{eq:doppler_0}.  After some algebra, we find:
\begin{equation}
  \label{eq:doppler_1}
  0 = \tilde{V} \bar{k}^{3} + \left( \cos{\theta{\scriptstyle_{kB}}} - \tilde{\omega}{\scriptstyle_{sc}} \right) \bar{k}^{2} + \tilde{V} \bar{k} - \tilde{\omega}{\scriptstyle_{sc}}
\end{equation}
where $\bar{k}$ $=$ kc/$\omega{\scriptstyle_{pe}}$, $\tilde{\omega}{\scriptstyle_{sc}}$ $=$ $\omega{\scriptstyle_{sc}}$/$\Omega{\scriptstyle_{ce}}$, $\tilde{V}$ $=$ V${\scriptstyle_{sw}}$$\cos{\theta{\scriptstyle_{kV}}}$/V${\scriptstyle_{Ae}}$, and V${\scriptstyle_{Ae}}$ $=$ B${\scriptstyle_{o}}$/($\mu{\scriptstyle_{o}}$n${\scriptstyle_{e}}$m${\scriptstyle_{e}}$)$^{1/2}$.  We already know $\theta{\scriptstyle_{kB}}$ and $\theta{\scriptstyle_{kV}}$ from our MVA results, thus we can solve Equation \ref{eq:doppler_1} for $\mid$\textbf{k}$\mid$ giving us our second wave number estimates.  \\
\indent  At very low frequencies ($\omega$ $\ll$ $\Omega{\scriptstyle_{ce}}$) and long wavelengths (kc/$\omega{\scriptstyle_{pe}}$ $\ll$ 1), this new equation reduces to the result found by \citet{coroniti82a} given by:
\begin{equation}
  \begin{split}
    \bar{k} = & \pm \left(\frac{ V{\scriptstyle_{sw}} \mid \cos{\theta{\scriptstyle_{kV}}} \mid }{ 2 V{\scriptstyle_{Ae}} \mid \cos{\theta{\scriptstyle_{kB}}} \mid }\right) + \\
    & \sqrt{ \left(\frac{ V{\scriptstyle_{sw}} \cos{\theta{\scriptstyle_{kV}}} }{ 2 V{\scriptstyle_{Ae}} \cos{\theta{\scriptstyle_{kB}}} }\right)^{2} + \frac{\omega{\scriptstyle_{sc}}}{\Omega{\scriptstyle_{ce}} \mid \cos{\theta{\scriptstyle_{kB}}} \mid } } \text{  .}  \label{eq:parawavenumber17}
  \end{split}
\end{equation}
Note that \citet{coroniti82a} used the absolute values to indicate that frequencies satisfying $\omega$ $>$ 0 imply $\omega{\scriptstyle_{sc}}$ $>$ 0.  The physical significance of the $\pm$ sign in Equation \ref{eq:parawavenumber17}, assuming $\mid$$\cos{\theta{\scriptstyle_{kV}}}$$\mid$ $=$ 1, is the following: $+$ sign corresponds to a $\omega$ red-shifted to the observed $\omega{\scriptstyle_{sc}}$; and the $-$ sign corresponds to a $\omega$ blue-shifted to the observed $\omega{\scriptstyle_{sc}}$.  Results from Equation \ref{eq:parawavenumber17} give us our third wave number estimates.  \\
\indent  Alternatively, the cold plasma dispersion relation for an electromagnetic lower hybrid wave is given by:
\begin{equation}
  \label{eq:lhws_em0}
  \left(\frac{ \omega }{ \Omega{\scriptstyle_{lh}} }\right)^{2} = \frac{1}{ 1 + \omega{\scriptstyle_{pe}}^{2}/k^{2}c^{2} } \left[ 1 + \frac{M{\scriptstyle_{i}}}{m{\scriptstyle_{e}}} \frac{ \cos^{2}{\theta{\scriptstyle_{kB}}} }{ 1 + \omega{\scriptstyle_{pe}}^{2}/k^{2}c^{2} } \right]
\end{equation}
where $\Omega{\scriptstyle_{lh}}$ is given by $\omega{\scriptstyle_{pi}}$/(1 + ($\omega{\scriptstyle_{pe}}$/$\Omega{\scriptstyle_{ce}}$)$^{2}$)$^{1/2}$.  Doppler shifting Equation \ref{eq:lhws_em0} and solving for $\mid$\textbf{k}$\mid$ gives us our fourth set of wave number estimates.  \\
\indent  Using the above equations and our observations, we will now present an example calculation using the waveform in Figure \ref{fig:3waveexamples}\textbf{C}.  We used two frequency filters on this wave at f${\scriptstyle_{sc}}$ $\sim$ 5-30 Hz and $\sim$ 120-200 Hz.  Thus, we separated Table \ref{tab:wavenums} into two sections by the different frequency filter ranges defined above.  The first part of each section of the table shows the background plasma parameters (B${\scriptstyle_{o}}$, n${\scriptstyle_{e}}$, and V${\scriptstyle_{sw}}$), filter range, and wave normal angles for this event.  Using the range of measured parameters shown in the first line of Table \ref{tab:wavenums}, we can estimate $\rho{\scriptstyle_{ce}}$ $\sim$ 1.6 km, c/$\omega{\scriptstyle_{pe}}$ $\sim$ 1.3 km, and f${\scriptstyle_{lh}}$ $\sim$ 10.3 Hz.  Notice that c/$\omega{\scriptstyle_{pe}}$ $\sim$ $\rho{\scriptstyle_{ce}}$, thus kc/$\omega{\scriptstyle_{pe}}$ will be $\sim$ k$\rho{\scriptstyle_{ce}}$ for this case.  Once we determine $\mid$\textbf{k}$\mid$, we can invert Equation \ref{eq:indexref1} to solve for $\omega$ using the solutions from Equations \ref{eq:indexref1}, \ref{eq:doppler_0}, and \ref{eq:parawavenumber17} to determine the wave phase speed, $\omega$/k.  The $\mid$\textbf{k}$\mid$-values determined from the Doppler shifted Equation \ref{eq:lhws_em0} were put back into Equation \ref{eq:lhws_em0} to find the corresponding $\omega$/k.  These values compose the last column in Table \ref{tab:wavenums}.


\clearpage

\begin{table}[htbp]
  \caption{Shock Parameters and Mach Number Ratios}
  \label{tab:machparams}
    \begin{tabular}{| c | c | c | c | c | c |}
      \hline
      \textbf{Date} & V${\scriptstyle_{shn}}$ (km/s) & $\theta{\scriptstyle_{Bn}}$ & M${\scriptstyle_{f}}$ & N${\scriptstyle_{i2}}$/N${\scriptstyle_{i1}}$ & M${\scriptstyle_{f}}$/M${\scriptstyle_{cr}}$   \\
      \hline
      1998-08-26  & 687 $\pm$ 27 & 82${\scriptstyle^{\circ}}$ $\pm$ 3${\scriptstyle^{\circ}}$ & 4.7 $\pm$ 0.2 & 2.9 $\pm$ 0.3  & 2.6 $\pm$ 0.4   \\
      1998-09-24  & 772 $\pm$ 96 & 82${\scriptstyle^{\circ}}$ $\pm$ 2${\scriptstyle^{\circ}}$ & 2.9 $\pm$ 0.1 & 2.2 $\pm$ 0.4  & 1.3 $\pm$ 0.2   \\
      2000-02-11  & 641 $\pm$ 13 & 87${\scriptstyle^{\circ}}$ $\pm$ 2${\scriptstyle^{\circ}}$ & 3.3 $\pm$ 0.1 & 3.3 $\pm$ 0.5  & 1.6 $\pm$ 0.2   \\
      2000-04-06  & 647 $\pm$ 98 & 70${\scriptstyle^{\circ}}$ $\pm$ 5${\scriptstyle^{\circ}}$ & 4.0 $\pm$ 0.6 & 3.8 $\pm$ 1.3  & 1.7 $\pm$ 0.3   \\
      \hline
    \end{tabular}
\end{table}

\begin{table}[htbp]
  \caption{Summary of Wave Observations}
  \label{tab:wavestats}
    \begin{tabular}{| c | c | c | c |}
      \hline \hline
      Wave Type & \# of TDSS & \# of wave & $\delta$B/B${\scriptstyle_{o}}$  \\
      & Events & vectors & (filtered) \\
      \hline
      LHWs & 23 & 118 & $\sim$0.054 $\pm$ 0.003 \\
      WWs  & 13 & 138 & $\sim$0.039 $\pm$ 0.002 \\
      MIXED  f${\scriptstyle_{low}}$  & 11 &  43 & $\sim$0.063 $\pm$ 0.011 \\
      MIXED  f${\scriptstyle_{high}}$ & 11 & 162 & $\sim$0.015 $\pm$ 0.001 \\
      \hline
      Total & 47 & 461 & \\
      \hline \hline
    \end{tabular}
\end{table}

\begin{table}[htb]
  \caption{Plasma parameters and Wave number estimates for Figure \ref{fig:3waveexamples}\textbf{C}}
  \label{tab:wavenums}
    \begin{tabular}{| c | c | c | c | c | c |}
      \hline  \hline
      \multicolumn{6}{|c|}{\textbf{Range of Parameters used for Doppler Shift Calculations}} \\
      \hline
      B${\scriptstyle_{o}}$ (nT) & n${\scriptstyle_{e}}$ (cm$^{-3}$) & V${\scriptstyle_{sw}}$ (km/s) & f${\scriptstyle_{sc}}$ (Hz) & $\theta{\scriptstyle_{kB}}$ (deg) & $\theta{\scriptstyle_{kV}}$ (deg) \\
      \hline
      $\sim$15.7 & $\sim$16.0 & $\sim$586 & 5-30 & 56$^{\circ}$-57$^{\circ}$ & 29$^{\circ}$-32$^{\circ}$ \\
      \hline
      \multicolumn{6}{|c|}{\textbf{Wave Numbers from Doppler Shift Calculations}} \\
      \hline
      Method/Equation & kc/$\omega{\scriptstyle_{pe}}$ & k (km$^{-1}$) & k${\scriptstyle_{\parallel}}$$\rho{\scriptstyle_{ce}}$ & k${\scriptstyle_{\perp}}$$\rho{\scriptstyle_{ce}}$ & $\omega$/k (km/s) \\
      \hline
      Equation \ref{eq:indexref1}                     & 0.1-0.4 & 0.2-0.3 & 0.1-0.3 & 0.2-0.4 & 280-660  \\
      Equation \ref{eq:doppler_1}                     & 0.1-0.4 & 0.1-0.3 & 0.09-0.3 & 0.1-0.4 & 270-680  \\
      Equation \ref{eq:parawavenumber17} ($-$ sign)   & 0.06-0.4 & 0.05-0.3 & 0.04-0.2 & 0.07-0.3 & 130-630  \\
      Equation  \ref{eq:parawavenumber17} ($+$ sign)  & 0.3-0.5 & 0.2-0.4 & 0.2-0.3 & 0.3-0.5 & 570-820  \\
      Doppler shifted Equation \ref{eq:lhws_em0}      & 0.1-0.4 & 0.1-0.3 & 0.09-0.3 & 0.1-0.4 & 280-680  \\
      \hline
      \multicolumn{6}{|c|}{\textbf{Range of Parameters used for Doppler Shift Calculations}} \\
      \hline
      B${\scriptstyle_{o}}$ (nT) & n${\scriptstyle_{e}}$ (cm$^{-3}$) & V${\scriptstyle_{sw}}$ (km/s) & f${\scriptstyle_{sc}}$ (Hz) & $\theta{\scriptstyle_{kB}}$ (deg) & $\theta{\scriptstyle_{kV}}$ (deg) \\
      \hline
      $\sim$15.7 & $\sim$16.0 & $\sim$586 & 120-200 & 15$^{\circ}$-25$^{\circ}$ & 82$^{\circ}$-89$^{\circ}$ \\
      \hline \hline
      \multicolumn{6}{|c|}{\textbf{Wave Numbers from Doppler Shift Calculations}} \\
      \hline
      Method/Equation & kc/$\omega{\scriptstyle_{pe}}$ & k (km$^{-1}$) & k${\scriptstyle_{\parallel}}$$\rho{\scriptstyle_{ce}}$ & k${\scriptstyle_{\perp}}$$\rho{\scriptstyle_{ce}}$ & $\omega$/k (km/s) \\ 
      \hline
      Equation \ref{eq:indexref1}                     & 0.6-1.0 & 0.5-0.8 & 0.7-1.1 & 0.2-0.5 & 1500-1780  \\
      Equation \ref{eq:doppler_1}                     & 0.6-1.0 & 0.5-0.8 & 0.7-1.2 & 0.2-0.5 & 1480-1770  \\
      Equation \ref{eq:parawavenumber17} ($-$ sign)   & 0.5-0.7 & 0.4-0.5 & 0.6-0.8 & 0.2-0.4 & 1360-1670  \\
      Equation  \ref{eq:parawavenumber17} ($+$ sign)  & 0.5-0.7 & 0.4-0.5 & 0.6-0.8 & 0.2-0.4 & 1380-1680  \\
      Doppler shifted Equation \ref{eq:lhws_em0}      & 0.5-1.0 & 0.4-0.8 & 0.5-1.2 & 0.2-0.5 & 1280-1780  \\
      \hline  \hline
    \end{tabular}
\end{table}

\begin{table}[htb]
  \caption{Wave parameters estimates for each wave mode}
  \label{tab:waveparams}
    \begin{tabular}{| c | c | c | c | c | c |}
      \hline  \hline
      Equation/Wave Type & kc/$\omega{\scriptstyle_{pe}}$ & k$\rho{\scriptstyle_{ce}}$ & k${\scriptstyle_{\parallel}}$$\rho{\scriptstyle_{ce}}$ & k${\scriptstyle_{\perp}}$$\rho{\scriptstyle_{ce}}$ & $\omega$/$\Omega$ \\
      \hline
      Doppler shifted Equation \ref{eq:lhws_em0}          & 0.02 - 5.0      & 0.01 - 5.0      & 0.0001 - 3.8    & 0.003 - 4.6     & 0.09 - 5.0  \\
      for LHWs ($\Omega$ $=$ $\omega{\scriptstyle_{lh}}$) & 0.50 $\pm$ 0.08 & 0.42 $\pm$ 0.07 & 0.17 $\pm$ 0.04 & 0.37 $\pm$ 0.06 & 1.94 $\pm$ 0.17  \\
      \hline
      Equation \ref{eq:doppler_1}                         & 0.2 - 1.0       & 0.2 - 0.9       & 0.2 - 0.8       & 0.003 - 1.7     & 0.03 - 0.9  \\
      for WWs ($\Omega$ $=$ $\Omega{\scriptstyle_{ce}}$)  & 0.48 $\pm$ 0.02 & 0.44 $\pm$ 0.02 & 0.40 $\pm$ 0.02 & 0.15 $\pm$ 0.01 & 0.18 $\pm$ 0.01  \\
      \hline
      Doppler shifted Equation \ref{eq:lhws_em0}                                   & 0.03 - 4.9      & 0.01 - 5.0      & 0.0003 - 4.1    & 0.001 - 4.3     & 0.07 - 5.0  \\
      for MIXED f${\scriptstyle_{low}}$ ($\Omega$ $=$ $\omega{\scriptstyle_{lh}}$) & 0.48 $\pm$ 0.11 & 0.39 $\pm$ 0.11 & 0.15 $\pm$ 0.06 & 0.35 $\pm$ 0.09 & 2.01 $\pm$ 0.25  \\
      \hline
      Equation \ref{eq:doppler_1} for MIXED                               & 0.1 - 4.9       & 0.07 - 4.9      & 0.003 - 4.5     & 0.005 - 5.0     & 0.003 - 0.96  \\
      f${\scriptstyle_{high}}$ ($\Omega$ $=$ $\Omega{\scriptstyle_{ce}}$) & 0.92 $\pm$ 0.06 & 0.71 $\pm$ 0.05 & 0.44 $\pm$ 0.04 & 0.50 $\pm$ 0.05 & 0.23 $\pm$ 0.01  \\
      \hline  \hline
    \end{tabular}
\end{table}

\begin{thebibliography}{57}
\providecommand{\natexlab}[1]{#1}
\expandafter\ifx\csname urlstyle\endcsname\relax
  \providecommand{\doi}[1]{doi:\discretionary{}{}{}#1}\else
  \providecommand{\doi}{doi:\discretionary{}{}{}\begingroup
  \urlstyle{rm}\Url}\fi

\bibitem[{\textit{{Akimoto} et~al.}(1985)\textit{{Akimoto}, {Papadopoulos}, and
  {Winske}}}]{akimoto85c}
{Akimoto}, K., K.~{Papadopoulos}, and D.~{Winske} (1985), {Lower-hybrid
  instabilities driven by an ion velocity ring}, \textit{J. Plasma Phys.},
  \textit{34}, 445--465, \doi{10.1017/S0022377800003007}.

\bibitem[{\textit{{Bougeret} et~al.}(1995)}]{bougeret95a}
{Bougeret}, J.-L., et~al. (1995), {Waves: The Radio and Plasma Wave
  Investigation on the Wind Spacecraft}, \textit{Space Sci. Rev.}, \textit{71},
  231--263, \doi{10.1007/BF00751331}.

\bibitem[{\textit{{Breneman} et~al.}(2010)\textit{{Breneman}, {Cattell},
  {Schreiner}, {Kersten}, {Wilson III}, {Kellogg}, {Goetz}, and
  {Jian}}}]{breneman10a}
{Breneman}, A., C.~{Cattell}, S.~{Schreiner}, K.~{Kersten}, L.~B. {Wilson III},
  P.~{Kellogg}, K.~{Goetz}, and L.~K. {Jian} (2010), {Observations of
  large-amplitude, narrowband whistlers at stream interaction regions},
  \textit{J. Geophys. Res.}, \textit{115}, A08,104, \doi{10.1029/2009JA014920}.

\bibitem[{\textit{{Brice}}(1964)}]{brice64a}
{Brice}, N. (1964), {Fundamentals of Very Low Frequency Emission Generation
  Mechanisms}, \textit{J. Geophys. Res.}, \textit{69}, 4515--+,
  \doi{10.1029/JZ069i021p04515}.

\bibitem[{\textit{{Cairns} and {McMillan}}(2005)}]{cairns05a}
{Cairns}, I.~H., and B.~F. {McMillan} (2005), {Electron acceleration by lower
  hybrid waves in magnetic reconnection regions}, \textit{Phys. Plasmas},
  \textit{12}, 102,110--+, \doi{10.1063/1.2080567}.

\bibitem[{\textit{{Cattell} et~al.}(1995)\textit{{Cattell}, {Wygant}, {Mozer},
  {Okada}, {Tsuruda}, {Kokubun}, and {Yamamoto}}}]{cattell95a}
{Cattell}, C., J.~{Wygant}, F.~S. {Mozer}, T.~{Okada}, K.~{Tsuruda},
  S.~{Kokubun}, and T.~{Yamamoto} (1995), {ISEE 1 and Geotail observations of
  low-frequency waves at the magnetopause}, \textit{J. Geophys. Res.},
  \textit{100}, 11,823--+, \doi{10.1029/94JA03146}.

\bibitem[{\textit{{Coroniti} et~al.}(1982)\textit{{Coroniti}, {Kennel},
  {Scarf}, and {Smith}}}]{coroniti82a}
{Coroniti}, F.~V., C.~F. {Kennel}, F.~L. {Scarf}, and E.~J. {Smith} (1982),
  {Whistler mode turbulence in the disturbed solar wind}, \textit{J. Geophys.
  Res.}, \textit{87}, 6029--6044, \doi{10.1029/JA087iA08p06029}.

\bibitem[{\textit{{Dyrud} and {Oppenheim}}(2006)}]{dyrud06}
{Dyrud}, L.~P., and M.~M. {Oppenheim} (2006), {Electron holes, ion waves, and
  anomalous resistivity in space plasmas}, \textit{J. Geophys. Res.},
  \textit{111}, 1302--+, \doi{10.1029/2004JA010482}.

\bibitem[{\textit{{Edmiston} and {Kennel}}(1984)}]{edmiston84}
{Edmiston}, J.~P., and C.~F. {Kennel} (1984), {A parametric survey of the first
  critical Mach number for a fast MHD shock.}, \textit{J. Plasma Phys.},
  \textit{32}, 429--441.

\bibitem[{\textit{{Feldman} et~al.}(1973)\textit{{Feldman}, {Asbridge}, {Bame},
  and {Montgomery}}}]{feldman73b}
{Feldman}, W.~C., J.~R. {Asbridge}, S.~J. {Bame}, and M.~D. {Montgomery}
  (1973), {Solar wind heat transport in the vicinity of the earth's bow
  shock.}, \textit{J. Geophys. Res.}, \textit{78}, 3697--3713,
  \doi{10.1029/JA078i019p03697}.

\bibitem[{\textit{{Gary} et~al.}(1994)\textit{{Gary}, {Scime}, {Phillips}, and
  {Feldman}}}]{gary94a}
{Gary}, S.~P., E.~E. {Scime}, J.~L. {Phillips}, and W.~C. {Feldman} (1994),
  {The whistler heat flux instability: Threshold conditions in the solar wind},
  \textit{J. Geophys. Res.}, \textit{99}, 23,391--+, \doi{10.1029/94JA02067}.

\bibitem[{\textit{{Gary} et~al.}(1999)\textit{{Gary}, {Skoug}, and
  {Daughton}}}]{gary99a}
{Gary}, S.~P., R.~M. {Skoug}, and W.~{Daughton} (1999), {Electron heat flux
  constraints in the solar wind.}, \textit{Phys. Plasmas}, \textit{6},
  2607--2612, \doi{10.1063/1.873532}.

\bibitem[{\textit{{Gurnett} and {Bhattacharjee}}(2005)}]{gurnett05}
{Gurnett}, D.~A., and A.~{Bhattacharjee} (2005), \textit{{Introduction to
  Plasma Physics: With Space and Laboratory Applications}}, ~ISBN
  0521364833.~Cambridge, UK: Cambridge University Press.

\bibitem[{\textit{{Hoppe} et~al.}(1981)\textit{{Hoppe}, {Russell}, {Frank},
  {Eastman}, and {Greenstadt}}}]{hoppe81a}
{Hoppe}, M.~M., C.~T. {Russell}, L.~A. {Frank}, T.~E. {Eastman}, and E.~W.
  {Greenstadt} (1981), {Upstream hydromagnetic waves and their association with
  backstreaming ion populations - ISEE 1 and 2 observations}, \textit{J.
  Geophys. Res.}, \textit{86}, 4471--4492, \doi{10.1029/JA086iA06p04471}.

\bibitem[{\textit{{Huba} et~al.}(1978)\textit{{Huba}, {Gladd}, and
  {Papadopoulos}}}]{huba78b}
{Huba}, J.~D., N.~T. {Gladd}, and K.~{Papadopoulos} (1978), {Lower-hybrid-drift
  wave turbulence in the distant magnetotail}, \textit{J. Geophys. Res.},
  \textit{83}, 5217--5226, \doi{10.1029/JA083iA11p05217}.

\bibitem[{\textit{{Kasper}}(2007)}]{kasper}
{Kasper}, J.~C. (2007), {Interplanetary Shock Database}, harvard-Smithsonian
  Center for Astrophysics, Online: \emph{http://www.cfa.harvard.edu/shocks/}.

\bibitem[{\textit{{Kasper} et~al.}(2006)\textit{{Kasper}, {Lazarus},
  {Steinberg}, {Ogilvie}, and {Szabo}}}]{kasper06a}
{Kasper}, J.~C., A.~J. {Lazarus}, J.~T. {Steinberg}, K.~W. {Ogilvie}, and
  A.~{Szabo} (2006), {Physics-based tests to identify the accuracy of solar
  wind ion measurements: A case study with the Wind Faraday Cups}, \textit{J.
  Geophys. Res.}, \textit{111}, 3105--+, \doi{10.1029/2005JA011442}.

\bibitem[{\textit{{Kennel} and {Petscheck}}(1966)}]{kennel66a}
{Kennel}, C.~F., and H.~E. {Petscheck} (1966), {Limit on stably trapped
  particle fluxes}, \textit{J. Geophys. Res.}, \textit{71}, 1--28.

\bibitem[{\textit{{Khrabrov} and {Sonnerup}}(1998)}]{khrabrov98}
{Khrabrov}, A.~V., and B.~U.~{\"O}. {Sonnerup} (1998), {Error estimates for
  minimum variance analysis}, \textit{J. Geophys. Res.}, \textit{103},
  6641--6652, \doi{10.1029/97JA03731}.

\bibitem[{\textit{{Krauss-Varban} and {Omidi}}(1991)}]{kraussvarban91}
{Krauss-Varban}, D., and N.~{Omidi} (1991), {Structure of medium Mach number
  quasi-parallel shocks - Upstream and downstream waves}, \textit{J. Geophys.
  Res.}, \textit{96}, 17,715--+, \doi{10.1029/91JA01545}.

\bibitem[{\textit{{Lemons} and {Gary}}(1978)}]{lemons78a}
{Lemons}, D.~S., and S.~P. {Gary} (1978), {Current-driven instabilities in a
  laminar perpendicular shock}, \textit{J. Geophys. Res.}, \textit{83},
  1625--1632, \doi{10.1029/JA083iA04p01625}.

\bibitem[{\textit{{Lengyel-Frey} et~al.}(1996)\textit{{Lengyel-Frey}, {Hess},
  {MacDowall}, {Stone}, {Lin}, {Balogh}, and {Forsyth}}}]{lengyelfrey96}
{Lengyel-Frey}, D., R.~A. {Hess}, R.~J. {MacDowall}, R.~G. {Stone}, N.~{Lin},
  A.~{Balogh}, and R.~{Forsyth} (1996), {Ulysses observations of whistler waves
  at interplanetary shocks and in the solar wind}, \textit{J. Geophys. Res.},
  \textit{101}, 27,555--27,564, \doi{10.1029/96JA00548}.

\bibitem[{\textit{{Lepping} et~al.}(1995)}]{lepping95}
{Lepping}, R.~P., et~al. (1995), {The Wind Magnetic Field Investigation},
  \textit{Space Sci. Rev.}, \textit{71}, 207--229, \doi{10.1007/BF00751330}.

\bibitem[{\textit{{Lin} et~al.}(1995)}]{lin95a}
{Lin}, R.~P., et~al. (1995), {A Three-Dimensional Plasma and Energetic Particle
  Investigation for the Wind Spacecraft}, \textit{Space Sci. Rev.},
  \textit{71}, 125--153, \doi{10.1007/BF00751328}.

\bibitem[{\textit{{Marsch} and {Chang}}(1983)}]{marsch83a}
{Marsch}, E., and T.~{Chang} (1983), {Electromagnetic lower hybrid waves in the
  solar wind}, \textit{J. Geophys. Res.}, \textit{88}, 6869--6880,
  \doi{10.1029/JA088iA09p06869}.

\bibitem[{\textit{{Masood} et~al.}(2006)\textit{{Masood}, {Schwartz},
  {Maksimovic}, and {Fazakerley}}}]{masood06a}
{Masood}, W., S.~J. {Schwartz}, M.~{Maksimovic}, and A.~N. {Fazakerley} (2006),
  {Electron velocity distribution and lion roars in the magnetosheath},
  \textit{Ann. Geophys.}, \textit{24}, 1725--1735,
  \doi{10.5194/angeo-24-1725-2006}.

\bibitem[{\textit{{Matsukiyo} and {Scholer}}(2006)}]{matsukiyo06b}
{Matsukiyo}, S., and M.~{Scholer} (2006), {On microinstabilities in the foot of
  high Mach number perpendicular shocks}, \textit{J. Geophys. Res.},
  \textit{111}, 6104--+, \doi{10.1029/2005JA011409}.

\bibitem[{\textit{{Mellott} and {Greenstadt}}(1988)}]{mellott88}
{Mellott}, M.~M., and E.~W. {Greenstadt} (1988), {Plasma waves in the range of
  the lower hybrid frequency - ISEE 1 and 2 observations at the earth's bow
  shock}, \textit{J. Geophys. Res.}, \textit{93}, 9695--9708,
  \doi{10.1029/JA093iA09p09695}.

\bibitem[{\textit{{Meyer-Vernet} and {Perche}}(1989)}]{meyervernet89a}
{Meyer-Vernet}, N., and C.~{Perche} (1989), {Tool kit for antennae and thermal
  noise near the plasma frequency}, \textit{J. Geophys. Res.}, \textit{94},
  2405--2415, \doi{10.1029/JA094iA03p02405}.

\bibitem[{\textit{{Moullard} et~al.}(1998)\textit{{Moullard}, {Burgess}, and
  {Bale}}}]{moullard98a}
{Moullard}, O., D.~{Burgess}, and S.~D. {Bale} (1998), {Whistler waves observed
  during an in-situ solar type III radio burst}, \textit{A{\&}A}, \textit{335},
  703--708.

\bibitem[{\textit{{Moullard} et~al.}(2001)\textit{{Moullard}, {Burgess},
  {Salem}, {Mangeney}, {Larson}, and {Bale}}}]{moullard01}
{Moullard}, O., D.~{Burgess}, C.~{Salem}, A.~{Mangeney}, D.~E. {Larson}, and
  S.~D. {Bale} (2001), {Whistler waves, Langmuir waves and single loss cone
  electron distributions inside a magnetic cloud: Observations}, \textit{J.
  Geophys. Res.}, \textit{106}, 8301--8314, \doi{10.1029/2000JA900144}.

\bibitem[{\textit{{Ogilvie} et~al.}(1995)}]{ogilvie95}
{Ogilvie}, K.~W., et~al. (1995), {SWE, A Comprehensive Plasma Instrument for
  the Wind Spacecraft}, \textit{Space Sci. Rev.}, \textit{71}, 55--77,
  \doi{10.1007/BF00751326}.

\bibitem[{\textit{{Russell} et~al.}(1969)\textit{{Russell}, {Holzer}, and
  {Smith}}}]{russell69a}
{Russell}, C.~T., R.~E. {Holzer}, and E.~J. {Smith} (1969), {OGO 3 observations
  of ELF noise in the magnetosphere. 1. Spatial extent and frequency of
  occurrence.}, \textit{J. Geophys. Res.}, \textit{74}, 755--777,
  \doi{10.1029/JA074i003p00755}.

\bibitem[{\textit{{Russell} et~al.}(1983)\textit{{Russell}, {Smith},
  {Tsurutani}, {Gosling}, and {Bame}}}]{russell83c}
{Russell}, C.~T., E.~J. {Smith}, B.~T. {Tsurutani}, J.~T. {Gosling}, and S.~J.
  {Bame} (1983), {Multiple spacecraft observations of interplanetary shocks:
  Characteristics of the upstream ULF turbulence}, \textit{NASA Conference
  Publication}, \textit{228}, 385--400.

\bibitem[{\textit{{Saito} and {Gary}}(2007)}]{saito07}
{Saito}, S., and S.~P. {Gary} (2007), {Whistler scattering of suprathermal
  electrons in the solar wind: Particle-in-cell simulations}, \textit{J.
  Geophys. Res.}, \textit{112}, 6116--+, \doi{10.1029/2006JA012216}.

\bibitem[{\textit{{Saito} et~al.}(2008)\textit{{Saito}, {Gary}, {Li}, and
  {Narita}}}]{saito08}
{Saito}, S., S.~P. {Gary}, H.~{Li}, and Y.~{Narita} (2008), {Whistler
  turbulence: Particle-in-cell simulations}, \textit{Phys. Plasmas},
  \textit{15}, 102,305--+, \doi{10.1063/1.2997339}.

\bibitem[{\textit{{Salem} et~al.}(2003)\textit{{Salem}, {Hubert}, {Lacombe},
  {Bale}, {Mangeney}, {Larson}, and {Lin}}}]{salem03}
{Salem}, C., D.~{Hubert}, C.~{Lacombe}, S.~D. {Bale}, A.~{Mangeney}, D.~E.
  {Larson}, and R.~P. {Lin} (2003), {Electron Properties and Coulomb Collisions
  in the Solar Wind at 1 AU: Wind Observations}, \textit{Astrophys. J.},
  \textit{585}, 1147--1157, \doi{10.1086/346185}.

\bibitem[{\textit{{Savoini} and {Lemb\`{e}ge}}(1995)}]{savoini95a}
{Savoini}, P., and B.~{Lemb\`{e}ge} (1995), {Heating and acceleration of
  electrons through the whistler precursor in 1-D and 2-D oblique shocks},
  \textit{Adv. Space Res.}, \textit{15}, 235--238,
  \doi{10.1016/0273-1177(94)00103-8}.

\bibitem[{\textit{{Shapiro} et~al.}(1993)\textit{{Shapiro}, {Shevchenko},
  {Solov'ev}, {Kalinin}, {Bingham}, {Sagdeev}, {Ashour-Abdalla}, {Dawson}, and
  {Su}}}]{shapiro93a}
{Shapiro}, V.~D., V.~I. {Shevchenko}, G.~I. {Solov'ev}, V.~P. {Kalinin},
  R.~{Bingham}, R.~Z. {Sagdeev}, M.~{Ashour-Abdalla}, J.~{Dawson}, and J.~J.
  {Su} (1993), {Wave collapse at the lower-hybrid resonance}, \textit{Phys.
  Fluids B}, \textit{5}, 3148--3162, \doi{10.1063/1.860652}.

\bibitem[{\textit{{Silin} et~al.}(2005)\textit{{Silin}, {B{\"u}chner}, and
  {Vaivads}}}]{silin05a}
{Silin}, I., J.~{B{\"u}chner}, and A.~{Vaivads} (2005), {Anomalous resistivity
  due to nonlinear lower-hybrid drift waves}, \textit{Phys. Plasmas},
  \textit{12}(6), 062,902--+, \doi{10.1063/1.1927096}.

\bibitem[{\textit{{Smith} and {Tsurutani}}(1976)}]{smith76a}
{Smith}, E.~J., and B.~T. {Tsurutani} (1976), {Magnetosheath lion roars},
  \textit{J. Geophys. Res.}, \textit{81}, 2261--2266,
  \doi{10.1029/JA081i013p02261}.

\bibitem[{\textit{{\v{S}tver\'{a}k} et~al.}(2008)\textit{{\v{S}tver\'{a}k},
  {Tr\'{a}vn{\'{\i}}\v{c}ek}, {Maksimovic}, {Marsch}, {Fazakerley}, and
  {Scime}}}]{stverak08a}
{\v{S}tver\'{a}k}, v., P.~{Tr\'{a}vn{\'{\i}}\v{c}ek}, M.~{Maksimovic},
  E.~{Marsch}, A.~N. {Fazakerley}, and E.~E. {Scime} (2008), {Electron
  temperature anisotropy constraints in the solar wind}, \textit{J. Geophys.
  Res.}, \textit{113}, 3103--+, \doi{10.1029/2007JA012733}.

\bibitem[{\textit{{\v{S}tver\'{a}k} et~al.}(2009)\textit{{\v{S}tver\'{a}k},
  {Maksimovic}, {Tr\'{a}vn{\'{\i}}\v{c}ek}, {Marsch}, {Fazakerley}, and
  {Scime}}}]{stverak09a}
{\v{S}tver\'{a}k}, v., M.~{Maksimovic}, P.~M. {Tr\'{a}vn{\'{\i}}\v{c}ek},
  E.~{Marsch}, A.~N. {Fazakerley}, and E.~E. {Scime} (2009), {Radial evolution
  of nonthermal electron populations in the low-latitude solar wind: Helios,
  Cluster, and Ulysses Observations}, \textit{J. Geophys. Res.}, \textit{114},
  5104, \doi{10.1029/2008JA013883}.

\bibitem[{\textit{{Tidman} and {Krall}}(1971)}]{tidman71a}
{Tidman}, D.~A., and N.~A. {Krall} (1971), \textit{{Shock waves in
  collisionless plasmas}}.

\bibitem[{\textit{{Torrence} and {Compo}}(1998{\natexlab{a}})}]{torrence98b}
{Torrence}, C., and G.~P. {Compo} (1998{\natexlab{a}}), {Wavelet Analysis
  Software}, atmospheric and Oceanic Sciences, University of Colorado, Online:
  \emph{http://paos.colorado.edu/research/wavelets/}.

\bibitem[{\textit{{Torrence} and {Compo}}(1998{\natexlab{b}})}]{torrence98a}
{Torrence}, C., and G.~P. {Compo} (1998{\natexlab{b}}), {A Practical Guide to
  Wavelet Analysis.}, \textit{Bull. Amer. Meteor. Soc.}, \textit{79}, 61--78,
  \doi{10.1175/1520-0477(1998)079}.

\bibitem[{\textit{{Vocks} and {Mann}}(2003)}]{vocks03a}
{Vocks}, C., and G.~{Mann} (2003), {Generation of Suprathermal Electrons by
  Resonant Wave-Particle Interaction in the Solar Corona and Wind},
  \textit{Astrophys. J.}, \textit{593}, 1134--1145, \doi{10.1086/376682}.

\bibitem[{\textit{{Vocks} et~al.}(2005)\textit{{Vocks}, {Salem}, {Lin}, and
  {Mann}}}]{vocks05a}
{Vocks}, C., C.~{Salem}, R.~P. {Lin}, and G.~{Mann} (2005), {Electron Halo and
  Strahl Formation in the Solar Wind by Resonant Interaction with Whistler
  Waves}, \textit{Astrophys. J.}, \textit{627}, 540--549, \doi{10.1086/430119}.

\bibitem[{\textit{{Walker} et~al.}(2008)\textit{{Walker}, {Balikhin},
  {Alleyne}, {Hobara}, {Andr\'{e}}, and {Dunlop}}}]{walker08a}
{Walker}, S.~N., M.~A. {Balikhin}, H.~S.~C.~K. {Alleyne}, Y.~{Hobara},
  M.~{Andr\'{e}}, and M.~W. {Dunlop} (2008), {Lower hybrid waves at the shock
  front: a reassessment}, \textit{Ann. Geophys.}, \textit{26}, 699--707.

\bibitem[{\textit{{Wilson III} et~al.}(2009)\textit{{Wilson III}, {Cattell},
  {Kellogg}, {Goetz}, {Kersten}, {Kasper}, {Szabo}, and
  {Meziane}}}]{wilsoniii09a}
{Wilson III}, L.~B., C.~A. {Cattell}, P.~J. {Kellogg}, K.~{Goetz},
  K.~{Kersten}, J.~C. {Kasper}, A.~{Szabo}, and K.~{Meziane} (2009),
  {Low-frequency whistler waves and shocklets observed at quasi-perpendicular
  interplanetary shocks}, \textit{J. Geophys. Res.}, \textit{114}, 10,106--+,
  \doi{10.1029/2009JA014376}.

\bibitem[{\textit{{Wilson III} et~al.}(2010)\textit{{Wilson III}, {Cattell},
  {Kellogg}, {Goetz}, {Kersten}, {Kasper}, {Szabo}, and
  {Wilber}}}]{wilsoniii10a}
{Wilson III}, L.~B., C.~A. {Cattell}, P.~J. {Kellogg}, K.~{Goetz},
  K.~{Kersten}, J.~C. {Kasper}, A.~{Szabo}, and M.~{Wilber} (2010),
  {Large-amplitude electrostatic waves observed at a supercritical
  interplanetary shock}, \textit{J. Geophys. Res.}, \textit{115}, 12,104--+,
  \doi{10.1029/2010JA015332}.

\bibitem[{\textit{{Wilson III} et~al.}(2012)}]{wilsoniii12d}
{Wilson III}, L.~B., et~al. (2012), {Observations of Electromagnetic Whistler
  Precursors at Supercritical Interplanetary Shocks}, \textit{Geophys. Res.
  Lett.}, \textit{39}, L08,109--+, \doi{10.1029/2012GL051581}.

\bibitem[{\textit{{Wu} et~al.}(1983)\textit{{Wu}, {Winske}, {Papadopoulos},
  {Zhou}, {Tsai}, and {Guo}}}]{wu83a}
{Wu}, C.~S., D.~{Winske}, K.~{Papadopoulos}, Y.~M. {Zhou}, S.~T. {Tsai}, and
  S.~C. {Guo} (1983), {A kinetic cross-field streaming instability},
  \textit{Phys. Fluids}, \textit{26}, 1259--1267, \doi{10.1063/1.864285}.

\bibitem[{\textit{{Wu} et~al.}(1984)}]{wu84a}
{Wu}, C.~S., et~al. (1984), {Microinstabilities associated with a high Mach
  number, perpendicular bow shock}, \textit{Space Sci. Rev.}, \textit{37},
  63--109, \doi{10.1007/BF00213958}.

\bibitem[{\textit{{Wygant} et~al.}(1987)\textit{{Wygant}, {Bensadoun}, and
  {Mozer}}}]{wygant87}
{Wygant}, J.~R., M.~{Bensadoun}, and F.~S. {Mozer} (1987), {Electric field
  measurements at subcritical, oblique bow shock crossings}, \textit{J.
  Geophys. Res.}, \textit{92}, 11,109--11,121, \doi{10.1029/JA092iA10p11109}.

\bibitem[{\textit{{Zhang} and {Matsumoto}}(1998)}]{zhang98a}
{Zhang}, Y., and H.~{Matsumoto} (1998), {Magnetic noise bursts near the
  interplanetary shock associated with the coronal mass ejection event on
  February 21, 1994: The Geotail observations}, \textit{J. Geophys. Res.},
  \textit{103}, 20,561--20,580, \doi{10.1029/98JA01234}.

\bibitem[{\textit{{Zhang} et~al.}(1998)\textit{{Zhang}, {Matsumoto}, and
  {Kojima}}}]{zhang98c}
{Zhang}, Y., H.~{Matsumoto}, and H.~{Kojima} (1998), {Lion roars in the
  magnetosheath: The Geotail observations}, \textit{J. Geophys. Res.},
  \textit{103}, 4615--4626, \doi{10.1029/97JA02519}.

\end{thebibliography}
\end{document}